\documentclass[nonacm, sigplan, screen]{acmart}
\pdfoutput=1

\setcopyright{acmlicensed}
\acmPrice{15.00}
\acmDOI{10.1145/3385412.3385973}
\acmYear{2020}
\copyrightyear{2020}
\acmSubmissionID{pldi20main-p81-p}
\acmISBN{978-1-4503-7613-6/20/06}
\acmConference[PLDI '20]{Proceedings of the 41st ACM SIGPLAN International Conference on Programming Language Design and Implementation}{June 15--20, 2020}{London, UK}
\acmBooktitle{Proceedings of the 41st ACM SIGPLAN International Conference on Programming Language Design and Implementation (PLDI '20), June 15--20, 2020, London, UK}
\usepackage{booktabs}   %% For formal tables:
\usepackage{subcaption} %% For complex figures with subfigures/subcaptions
\usepackage[english]{babel}
\usepackage{pbox}
\usepackage{xspace}

\usepackage{amssymb}
\usepackage{stmaryrd}
\usepackage{bbm}
\usepackage{dblfloatfix}
\usepackage{sepfootnotes}

\newsymbolfootnotes[page]{a}

\usepackage{tikz}
\usetikzlibrary{litmus, litmus.listings}
\usetikzlibrary{decorations.pathreplacing}

\usetikzlibrary{external} %%include library
\usepackage{pgfplots}

\usepackage{enumitem}

\newcommand\figLeftMargin[0]{\addtolength{\leftskip}{7mm}}

\newcommand{\CMMT}{/\nolinebreak\hspace{-.18em}/}

\newcommand\void[1]{}
\newcommand\K[1]{\ensuremath{\textsf{\sf #1}}}
\newcommand\KK[1]{\ensuremath{\K{\textbf{#1}}}}

\newcommand\fld[1]{\textsf{#1}}
\newcommand\rel[1]{\fld{#1}}
\newcommand\func[1]{\rel{#1}}
\newcommand\type[1]{\ensuremath{\mathit{#1}}}
\newcommand\tcons[1]{\KK{#1}}
\newcommand\obj[1]{\textit{#1}}

\newcommand\evttype{\type{event}}
\newcommand\nattype{\type{nat}}
\newcommand\booltype{\type{bool}}
\newcommand\bytetype{\type{byte}}

\newcommand\candidatetype{\type{candidate\_execution}}

\newcommand\listtype[1]{\type{list\ #1}}
\newcommand\settype[1]{\type{set\ #1}}

\newcommand\definerel[3]{%
  \expandafter\newcommand\csname#1\endcsname{\rel{#2}\xspace}%
  \expandafter\newcommand\csname#1c\endcsname{\color{#3}}%
  \expandafter\newcommand\csname#1r\endcsname{\KK{\color{#3} #1}}%
}

\definecolor{coColor}    {rgb}{0.647,0.165,0.165} % svgname Brown
\definecolor{depColor}  {rgb}{0.294,0.000,0.510} % svgname Indigo
\definecolor{frColor}    {rgb}{1,0.648,0}         % svgname

\newcommand\rf{\rel{reads-from}\xspace}
\newcommand\rfc{\color{red}}
\newcommand\rfr{\KK{\rfc rf}}

\newcommand\rbf{\rel{reads-byte-from}\xspace}
\newcommand\rbfc{\rfc}
\newcommand\rbfr{\KK{\rbfc rbf}}

\newcommand\seqb{\rel{sequenced-before}\xspace}
\newcommand\seqbc{\color{black}}
\newcommand\seqbr{\KK{\seqbc sb}}

\newcommand\po{\rel{program-order}\xspace}

\newcommand\asw{\rel{additional-synchronizes-with}\xspace}
\newcommand\aswc{\color{gray}}
\newcommand\aswr{\KK{\aswc asw}}

\newcommand\hb{\rel{happens-before}\xspace}
\newcommand\hbc{\color{OliveGreen}}
\newcommand\hbr{\KK{\hbc hb}}

\newcommand\sw{\rel{synchronizes-with}\xspace}
\newcommand\swc{\color{purple}}
\newcommand\swr{\KK{\swc sw}}

\newcommand\tot{\rel{total-order}\xspace}
\newcommand\totc{\color{orange}}
\newcommand\totr{\KK{\totc tot}}

\definerel{addrP}{address-dependency-pre}{depColor}
\definerel{dataP}{data-dependency-pre}{depColor}
\definerel{ctrlP}{control-dependency-pre}{depColor}
\definerel{addr}{address-dependency}{depColor}
\definerel{data}{data-dependency}{depColor}
\definerel{ctrl}{control-dependency}{depColor}
\definerel{rmw}{read-modify-write}{depColor}
\definerel{aob}{atomic-ordered-before}{black}
\definerel{bob}{barrier-ordered-before}{black}
\definerel{dob}{dependency-ordered-before}{black}
\definerel{ob}{ordered-before}{black}
\definerel{iob}{initial-ordered-before}{black}
\definerel{sthd}{same-thread}{black}
\definerel{rfi}{reads-from-internal}{red}
\definerel{rfe}{reads-from-external}{red}
\definerel{fri}{from-reads-internal}{frColor}
\definerel{fre}{from-reads-external}{frColor}
\definerel{frb}{bytewise-from-reads}{frColor}
\definerel{coi}{coherence-internal}{coColor}
\definerel{coe}{coherence-external}{coColor}
\definerel{ca}{coherence-after}{coColor}
\definerel{obs}{observed}{black}
\definerel{polocb}{program-order-same-byte-location}{black}

\newcommand\ordun{\KK{Un}\xspace}
\newcommand\ordsc{\KK{SC}\xspace}
\newcommand\ordi{\KK{I}\xspace}

\newcommand\Wun{\K{$\text{W}_{\ordun}$}}
\newcommand\Wsc{\K{$\text{W}_{\ordsc}$}}

\newcommand\Wany{\K{$\text{W}_{\hspace{-0.2em}\textit{any}}$}}

\newcommand\Run{\K{$\text{R}_{\ordun}$}}
\newcommand\Rsc{\K{$\text{R}_{\ordsc}$}}

\newcommand\RMWsc{\K{$\text{RMW}_{\ordsc}$}}

\lstdefinelanguage{JavaScript}{
  sensitive=true,
  alsoletter={., -},
  morecomment=[l]{//},
  showstringspaces=false,
  morekeywords=[2]{if},
  morekeywords=[3]{A-.store, A-.load, Atomics.store,Atomics.load,Atomics.wait,Atomics.notify},
  basicstyle=\ttfamily,
  commentstyle=\color{black!75},
  identifierstyle=,
  keywordstyle=[2]\color{purple},
  keywordstyle=[3]\color{purple},
}

\litmusset{
  use listing={basicstyle=\normalsize\ttfamily},
  JavaScript/.style={
    /litmus/listing options/.prefix={language=JavaScript,},
    /litmus/base language,
    /litmus/sw/.litmus relation=purple,
    /litmus/tot/.litmus relation=orange,
    /litmus/hb/.litmus relation=OliveGreen,
    /litmus/sb/.litmus relation=black,
    /litmus/asw/.litmus relation=gray,
  },
  every litmus/.append style={
    node font=\normalsize,
    every thread header/.append style={draw=none, node font=\bfseries},
    every simple instructions/.append style={
      /tikz/row sep={12pt,between origins},
    },
    every event/.append style={outer sep=2pt},
    every relation/.append style=overlay,
  },
  every assem litmus/.append style={threads distance=1em},
  AArch64/.append style={%
    /litmus/listing options/.prefix={style=litmus,},
  },
  simple instructions without header/.style={
    /litmus/add thread header=false,
    /litmus/add thread header/.code={},
    /litmus/simple instructions,
  },
  threads baseline/.style={
    /litmus/every thread header/.append style={alias=thread-##1},
    baseline=(thread-0.base),
  },
}

\newcommand{\xxpair}[2]{\ensuremath{\left\langle #1, #2 \right\rangle}}
\newcommand{\xxtriple}[3]{\ensuremath{\left\langle #1, #2, #3 \right\rangle}}

\newcommand\blfootnote[1]{%
  \begingroup
  \renewcommand\thefootnote{}\footnote{#1}%
  \addtocounter{footnote}{-1}%
  \endgroup
}

\begin{document}

%% Title information
%\title[JS]{Untitled JavaScript Memory Model Paper}         %% [Short Title] is optional;
                                        %% when present, will be used in
                                        %% header instead of Full Title.
\title[Repairing and Mechanising the JavaScript Relaxed Memory Model]{Repairing and Mechanising the\\ JavaScript Relaxed Memory Model} % JP: my attempt at an alliteration
%\titlenote{with title note}             %% \titlenote is optional;
                                        %% can be repeated if necessary;
                                        %% contents suppressed with 'anonymous'
%\subtitle{Subtitle}                     %% \subtitle is optional
%\subtitlenote{with subtitle note}       %% \subtitlenote is optional;
                                        %% can be repeated if necessary;
                                        %% contents suppressed with 'anonymous'

%% Author information
%% Contents and number of authors suppressed with 'anonymous'.
%% Each author should be introduced by \author, followed by
%% \authornote (optional), \orcid (optional), \affiliation, and
%% \email.
%% An author may have multiple affiliations and/or emails; repeat the
%% appropriate command.
%% Many elements are not rendered, but should be provided for metadata
%% extraction tools.

%% Author with single affiliation.
\author[Watt]{Conrad Watt}
%\authornote{with author1 note}          %% \authornote is optional;
                                        %% can be repeated if necessary
%\orcid{nnnn-nnnn-nnnn-nnnn}             %% \orcid is optional

\affiliation{%
  \institution{University of Cambridge}
  \country{UK}}
  % \position{Position1}
  % \department{Department1}              %% \department is recommended
  % \institution{Institution1}            %% \institution is required
  % \streetaddress{Street1 Address1}
  % \city{City1}
  % \state{State1}
  % \postcode{Post-Code1}
%}
\email{conrad.watt@cl.cam.ac.uk}          %% \email is recommended

%% our author list on the top of the page is still too long when using
%% the initial of the first name, so just doing surnames in short
%% version

%% Author with two affiliations and emails.
\author[Pulte]{Christopher Pulte}
\affiliation{%
  \institution{University of Cambridge}
  \country{UK}}
\email{christopher.pulte@cl.cam.ac.uk}

\author[Podkopaev]{Anton Podkopaev}
\affiliation{%
  \institution{HSE / MPI-SWS}
  \country{Russia / Germany}}
\email{podkopaev@mpi-sws.org}

\author[Barbier]{Guillaume Barbier}
\affiliation{%
  \institution{ENS Rennes}
  \country{France}}
\email{guillaume.barbier@ens-rennes.fr}

\author[Dolan]{Stephen Dolan}
\affiliation{%
  \institution{University of Cambridge}
  \country{UK}}
\email{stephen.dolan@cl.cam.ac.uk}

\author[Flur]{Shaked Flur}
\affiliation{%
  \institution{University of Cambridge\textsuperscript{$\ast$}}
  \country{UK}}
\email{shaked.flur@cl.cam.ac.uk}

\author[Pichon-Pharabod]{Jean Pichon-Pharabod}
\affiliation{%
  \institution{University of Cambridge}
  \country{UK}}
\email{jean.pichon@cl.cam.ac.uk}

\author[Guo]{Shu-yu Guo}
\affiliation{%
  \institution{Bloomberg LP\textsuperscript{$\ast$}}
  \country{USA}}
\email{shu@rfrn.org}

%% Abstract
%% Note: \begin{abstract}...\end{abstract} environment must come
%% before \maketitle command
%!TEX root = main.tex

\begin{abstract}
Modern JavaScript includes the SharedArrayBuffer feature, which provides access to true shared memory concurrency.
SharedArrayBuffers are simple linear buffers of bytes, and the JavaScript specification defines an axiomatic relaxed memory model to describe their behaviour.
While this model is heavily based on the C/C++11 model, it diverges in some key areas.
JavaScript chooses to give a well-defined semantics to data-races, unlike the ``undefined behaviour'' of C/C++11.
Moreover, the JavaScript model is \textit{mixed-size}.
This means that its accesses are not to discrete locations, but to (possibly overlapping) ranges of bytes.

We show that the model, in violation of the design intention, does not support a compilation scheme to ARMv8 which is used in practice.
We propose a correction, which also incorporates a previously proposed
fix for a failure of the model to provide Sequential Consistency of Data-Race-Free
programs (SC-DRF), an important correctness condition.
We use model checking, in Alloy, to generate small counter-examples for these deficiencies, and investigate our correction.
%
% Am I correct to say "both" here? Or is the model-checked property the proper SC-DRF and the Coq-proved property the model-internal one?
%
To accomplish this, we also develop a mixed-size extension to the existing ARMv8 axiomatic model.
% dropped the ``experimentally validate'' part, think it'll sound weaker than it needs to, as though we didn't have confidence in that model}

Guided by our Alloy experimentation, we mechanise (in Coq) the JavaScript model (corrected and uncorrected), our ARMv8 model, and, for the corrected JavaScript model, a ``model-internal'' SC-DRF proof and a compilation scheme correctness proof to ARMv8.
In addition, we investigate a non-mixed-size subset of the corrected JavaScript model, and give proofs of compilation correctness for this subset to x86-TSO, Power, RISC-V, ARMv7, and (again) ARMv8, via the Intermediate Memory Model (IMM).  

As a result of our work, the JavaScript standards body (ECMA TC39) will include fixes for both issues in an upcoming edition of the specification.

\end{abstract}



%% Keywords
%% comma separated list
  %% \keywords are mandatory in final camera-ready submission

%% \maketitle
%% Note: \maketitle command must come after title commands, author
%% commands, abstract environment, Computing Classification System
%% environment and commands, and keywords command.
\maketitle

\blfootnote{\textsuperscript{$\ast$} At the time the work was done.}

%!TEX root = main.tex

\section{Introduction}

JavaScript is widely publicised as a ``single-threaded language''~\cite{js-event-loop}, with asynchronously dispatched events processed by a single event loop.
JavaScript (JS) allows the use of threads, called \textit{Web Workers}, for parallel computation, but until recently these were not allowed to share access to mutable state, and inter-thread communication was restricted purely to message-passing~\cite{boduch2015javascript}.
However, a new feature of JavaScript, SharedArrayBuffer, allows true concurrent access to a low-level shared resource~\cite{ecmascript_sharedarraybuffer}.
SharedArrayBuffers are simple linear mutable byte buffers, and, unlike other JavaScript objects, references to the same SharedArrayBuffer may be held by multiple Web Workers simultaneously.

SharedArrayBuffers were originally specified and implemented by all major browsers in 2017, but were disabled shortly after due to concerns about Spectre and Meltdown.
Now that mitigations have been developed, the feature has been re-introduced into the %Web
web ecosystem~\cite{chromium_sab}.

SharedArrayBuffers have several important uses.
They are the only mechanism in JavaScript for true shared-memory concurrency.
C++ is often compiled to asm.js~\cite{asmjs:2014}, a fast JavaScript subset, for use on the web, and concurrent C++ objects must be allocated on SharedArrayBuffers in the compiled program.
SharedArrayBuffers also provide the mechanism by which JavaScript may interoperate with concurrent \mbox{WebAssembly} programs~\cite{Wasm:2017}.

The JavaScript specification must define the concurrent
behaviours that are possible when the same SharedArrayBuffer is
accessed concurrently by multiple Web Workers; this is done through a \textit{relaxed memory model}.
The relaxed memory model of JavaScript was
designed to conform to an ambitious set of
requirements~\cite{javascript-sharedmem-rationale}:

\paragraph{Mixed-size accesses} JavaScript's concurrency is
  \textit{mixed-size}, in the sense of
  \citet{Flur:2017:MCA:3009837.3009839}: accesses
  are not to individual, discrete locations, but instead to ranges of
  byte locations, which may overlap with each other.

\paragraph{Mixed atomic and non-atomic accesses} JavaScript has no concept
  of an ``atomic location'', so atomic and non-atomic accesses may be
  arbitrarily combined on the same location. JavaScript has only
  one type of atomic access, \KK{SeqCst}, while C/C++11 also has so-called ``low-level atomics'' such as release/acquire.

\paragraph{No undefined behaviour} The JavaScript language does not have a
  concept of undefined behaviour, so all programs must have behaviour
  defined by the standard. This remains true even in the presence of
  data-races, although the defined behaviour is then extremely weak.

\paragraph{C++-compatible compilation} JavaScript atomic accesses are to
  use the same compilation scheme as C++ SC-atomic accesses~\cite[p. 17]{javascript-sharedmem-rationale}.

\paragraph{SC-DRF} Programs that are free of data races must have
  sequentially consistent
  behaviour~\cite[p. 8]{javascript-sharedmem-rationale}.\\
 % \todojp{Although we argue in \S\ref{sec:scdrf} that the JavaScript model uses too weak a definition.}

Whatever JavaScript's general reputation, it should be emphasised that its current specification is particularly rigorous.
Because of its ubiquity on the Web %is it always capitlised?
and the large number of language implementers, great care is taken to ensure that its features are precisely defined.\textsuperscript{$1$}
In particular, its relaxed memory model is defined using an unambiguous, semi-formal pseudocode, which takes inspiration from previous formalisations of the memory models of C/C++11~\cite{Batty:2011} and Java~\cite{Manson:2005:JMM:1040305.1040336}.
In addition, SharedArrayBuffers have been carefully designed to
participate as little as possible in JavaScript's complicated object inheritance model, effectively allowing us
to reason about them in isolation.
\blfootnote{\textsuperscript{$1$} JavaScript's least intuitive behaviours arise not out of failures of its current specification process, but out of a requirement to be backwards-compatible with earlier versions of the language. %
Often, this means specifying a strange behaviour for legacy reasons.}
\setcounter{footnote}{1}

\pagebreak

\subsection{Mixed-Size}
\label{sec:intro-mixed-size}

In some respects, the JavaScript relaxed memory model is similar to that of C/C++~\cite{Batty:2011}, sharing an axiomatic nature and several core definitions and relations.
However, the mixed-size nature of the JavaScript model is a substantial
complication in its verification.
Unlike what prior work typically assumes, memory accesses can be of different sizes (byte-widths); hence, two accesses may overlap without having the exact
same ``footprint'', adding another dimension of complexity to the model, and limiting our ability to make use of prior work in which this is assumed not to occur.

To the best of our knowledge, only a single previous work, \citet{Flur:2017:MCA:3009837.3009839}, deals with the formal verification of compilation of a mixed-size relaxed memory model.
%
% \citet{Flur:2017:MCA:3009837.3009839} %
%
Flur et al. concentrate mainly on architecture-level mixed-size behaviours, proposing operational mixed-size models for the ARMv8 and POWER architectures.
Their work describes an extension to an existing formalisation of C/C++11, adding mixed-size non-atomics, and gives a sketch hand-proof that the resulting model can be correctly compiled to POWER, acknowledging that fuller verification is an open problem. % \todocp{And I think this is also mixed-sizing the POWER hand-proof that was later found to be incorrect? We should check.}
Even this mixed-size C/C++11 model only allows mixed-size non-atomics,
on which a data race leads to undefined behaviour. The JavaScript memory model, in contrast, must give well-defined behaviour even in the case of data races between partially-overlapping accesses.

Two other papers deal with mixed-size models.
\citet{weakeningwasm} describe a memory model for WebAssembly that is closely related to the JavaScript memory model.
The authors do not attempt a proof of correctness of compilation, again declaring it as an open problem.
They report on a deficiency in the JavaScript memory model which we investigate further as part of this work.
Finally, the EMME tool~\cite{emmejs} represents an earlier investigation of JavaScript's memory model using the Alloy model checker~\cite{Jackson:2002}.
This tool is engineered primarily as a test oracle, and the work does not attempt compilation scheme verification.
We also make use of Alloy during this work, but primarily for compilation scheme investigation, in the style of Memalloy~\cite{Wickerson:2017:ACM:3009837.3009838}.
We discuss differences in our approaches in~\S\ref{sec:related}.

\subsection{Our Contributions}
\paragraph{ARMv8 compilation scheme failure}
We have discovered that, on ARMv8, compiling JavaScript atomics by following the standard
compilation scheme for C++ SC atomics allows behaviours which violate the guarantees of the JavaScript memory model.
This compilation scheme is already implemented in Google's V8 JavaScript engine~\cite{v8-rel-acq}, and we are able to
observe the violating behaviour experimentally through Web browsers on
real hardware.
After consultation with implementers and ECMA TC39, the JavaScript standards body, we concluded that the JavaScript memory model should be weakened in order to support this scheme (\S\ref{sec:arm_compilation_failure}).

\pagebreak

\paragraph{Fixing the JS specification}
The JavaScript model also fails to guarantee Sequential
  Consistency for Data-Race-Free programs (SC-DRF), a crucial
  correctness condition~\cite{Adve:1990:WON:325164.325100,Lahav:2017:RSC:3062341.3062352,Boehm:2008,Gharachorloo92programmingfor}. This was discovered in previous
  work~\cite{weakeningwasm}, which proposed a strengthening of the
  model to restore SC-DRF, but did not verify that the
  strengthened condition is supported by existing compilation schemes.
  We integrate our fix for the ARMv8 compilation issue with this previous proposal, obtaining a combined fix to the JavaScript model (\S\ref{sec:scdrf}).
On the strength of our model checking and verification work (detailed below), this combined fix was adopted by the standards body for inclusion in an upcoming edition of the standard.

\paragraph{Alloy model checking}
Using the Alloy model checker~\cite{Jackson:2002}, we can automatically
  find counter-examples exemplifying the above two deficiencies,
  following the approach of Memalloy~\cite{Wickerson:2017:ACM:3009837.3009838}, but here, for the first time, applied in a mixed-size context.
  We also use Alloy to inform our Coq compilation scheme verification of the revised model, by %
  %% first checking up to a bound that possible approaches for constructing the full proof admit no counter-examples.
  essentially model checking (up to a bound) the main construction used by that proof (of JavaScript-allowed executions from ARMv8-allowed executions).

\paragraph{Mixed-size axiomatic ARMv8 model}
To enable verifying the compilation scheme, we define, first in
  Alloy, then in Coq, a novel mixed-size ARMv8 axiomatic model, as a generalisation
  of ARM's axiomatic reference model, and validate it with respect to
  Flat~\cite{Flur:2017:MCA:3009837.3009839,PulteEtAl2018}, a
  well-tested mixed-size operational model for ARMv8
  (\S\ref{arm_mixed}).

\paragraph{Coq proofs}
We mechanise the JavaScript and mixed-size ARMv8 models in Coq, and give a proof of compilation scheme correctness, and the ``model-internal'' SC-DRF property (see \S\ref{sec:scdrf}) for the revised model.
We investigate the (subtle) circumstances under which the mixed-size model may be reduced to an equivalent non-mixed-size (hereafter ``uni-size'') model.
We define a
  uni-size subset of the JavaScript model and prove, in Coq,
  compilation scheme correctness for this subset to x86-TSO, Power,
  RISC-V, ARMv7, and (again) ARMv8 via the %Intermediate Memory Model
  IMM
  model~\cite{Podkopaev:2019:BGP:3302515.3290382,Moiseenko-al:CoRR19}
  (\S\ref{sec:coq}).

\paragraph{Thread suspension specification}
Looking beyond memory accesses, JavaScript also defines thread suspension operations: ``\texttt{Atomics.wait}'', conditionally blocking a thread, and ``\texttt{Atomics.notify}'', unblocking waiting threads.
The synchronization guarantees of these operations were not integrated into the formal model, leading to ambiguities which we correct (\S\ref{sec:wait-notify}).
\\

\noindent Our artefacts are distributed as supplemental material~\cite{papersuppl}.

\pagebreak

\subsection{Non-contributions}

C/C++11-style ``out-of-thin-air'' executions~\cite{10.1007/978-3-662-46669-8_12} are admitted by the JavaScript model for certain programs with racing non-atomics~\cite{weakeningwasm}.
We acknowledge this is a deficiency of the model, but we do not attempt to solve this long-standing problem here; proposed solutions for C/C++11 involve either performance sacrifices~\cite{Lahav:2017:RSC:3062341.3062352} or the adoption of a radically different model~\cite{10.1145/3009837.3009850}.

%!TEX root = main.tex

\section{JavaScript's Shared Memory}

\label{sec:backgroundjs}

%!TEX root = ../main.tex
%
\begin{figure*}[h]
\begin{minipage}{0.5\textwidth}
  \begin{tikzpicture}[%
    /litmus/append to tikz path search,
    kind=assem,
    JavaScript,
    threads baseline,
  ]
  % Thread 0
  \node[simple instructions] {
    |x[0] = 3;  | \\
    |Atomics.store(x,1,5);  | \\
  };

  % Thread 1
  \node[simple instructions] {
    |r0 = Atomics.load(x,1);| \\
    |if (r0 == 5) {| \\
    |  r1 = x[0];             | \\
        |}| \\
  };
  
   \node at (2.5,1.2)   (x) {\texttt{x = new Int32Array(new SharedArrayBuffer[1024]);}};
  \end{tikzpicture}%
\caption{A simple JavaScript program.}%
\label{fig:MP-simple}%
\end{minipage}%\hspace{0.05\textwidth}%
\begin{minipage}{0.5\textwidth}
\centering
%\vspace{-0.22cm}
\begin{tikzpicture}[%
  /litmus/append to tikz path search,
  kind=events,
  JavaScript,
  threads baseline,
]
% Thread 0
\node[simple instructions] {
  a:W\textsubscript{Un} x[0..3]=3\hspace{1cm} \\
  b:W\textsubscript{SC} x[4..7]=5 \\
};

% Thread 1
\node[simple instructions] {
  c:R\textsubscript{SC} x[4..7]=5 \\
  \\[-0.1cm]
  d:R\textsubscript{Un} x[0..3]=3 \\
};

\node at (1.8,1.3)   (x) {W\textsubscript{I} b[0..1024]=0};

% [rf, out=-165, in=175, looseness=1.45, swap, label pos=0.75]
% disable edge routing
\draw
      (x) edge[hb, swap, in=15, out=-90] (a)
      (b) edge[disable edge routing, multi'={sw,rf,hb}, color=sw color, swap, label pos=0.2] (c)
      (a) edge[rf, out=-177, in=190, looseness=1.6, label pos=0.75] (d)
      (x) edge[hb, in=165, out=-90] (c);
 
\path[use as bounding box] (-2.65,-0.95) rectangle (c.east |- x.north); % adjust to fit

\end{tikzpicture}
\vspace{0.58\baselineskip}
\caption{A candidate execution for Fig.~\ref{fig:MP-simple}.}%
\label{fig:MP-simple-exec}%
\end{minipage}
\end{figure*}

As previously mentioned, JavaScript allows threads to concurrently access SharedArrayBuffer objects, which are simply zero-initialised raw buffers of bytes.

To access a SharedArrayBuffer, the programmer must declare a special wrapper around it called a \textit{typed array}.
Every typed array has a \textit{width}, which is the number of consecutive bytes in the underlying SharedArrayBuffer that a single load or store on the typed array will access.
Fig.~\ref{fig:MP-simple} depicts a simple two-threaded program that initially declares a single SharedArrayBuffer of 1024 bytes, and wraps it in a typed array with a width of 32 bits (4 bytes).
The two threads then perform a simple \textit{message-passing} procedure, which we use to illustrate JavaScript's two access modes.
Thread 0 writes the value 1 to location 0 (the \textit{message}) using a standard non-atomic access (note that this corresponds to bytes 0-3 of the underlying SharedArrayBuffer). 
It then writes 1 to location 1 (the \textit{flag}, at bytes 4-7) using a sequentially consistent atomic access.
Thread 1 reads location 1 atomically and then, only if it observes thread 0's write, reads location 0.

\begin{sloppypar}
The paired atomic read/write on location 1 give strong ordering guarantees.
Two possible outcomes are allowed: either \mbox{\texttt{r0 = 5 $\wedge$ r1 = 3}} or \mbox{\texttt{r0 = 0}}.
In particular, the outcome \mbox{\texttt{r0 = 5 $\wedge$ r1 = 0}}, where the flag is observed as set but the message is not received, is not allowed.
However, if either of the two atomic operations are replaced with non-atomics, then the outcome \mbox{\texttt{r0 = 5 $\wedge$ r1 = 0}} \textit{can} be experimentally observed.
This is an example of \textit{relaxed memory behaviour}.
\end{sloppypar}

As a convention, when we depict JavaScript code fragments, all accesses
will be assumed to be to 32-bit typed arrays unless otherwise
stated.
A single SharedArrayBuffer may be wrapped by multiple typed arrays of different widths, leading to mixed-size behaviours, where accesses partially overlap with each other.

JavaScript also provides a low-level mechanism for manipulating SharedArrayBuffers called a DataView.
DataViews only offer non-atomic operations, which may uniquely be \textit{unaligned}.
DataViews are far less commonly-used than typed arrays; they have historically been avoided due to performance problems~\cite{v8dataview}, with Emscripten~\cite{10.1145/2048147.2048224} (a key JavaScript-producing toolchain) generating code that uses typed arrays exclusively.

Most of our results cover DavaView-generated unaligned accesses, with the exception of our Coq compilation scheme correctness proof (\S\ref{sec:coq-arm}), which handles only the aligned (but still possibly mixed-size) accesses generated by typed arrays.

We now define the JavaScript relaxed memory model as it appears in the latest (10\textsuperscript{th}) edition of the specification~\cite{ecmascript_mm}.
This model describes the range of relaxed memory behaviours that a JavaScript program is allowed to exhibit.
In the course of this paper, we will present and discuss alterations to the model that have been accepted for inclusion in a future edition, which fix several deficiencies in the model as presented here.

\subsection{Thread-Local and Axiomatic Semantics}
JavaScript, like C++, has an axiomatic relaxed memory model.
%declares
This means that the allowed concurrent behaviours are stated as whole-execution axiomatic constraints.
A program's semantics is defined in two layers.
First, an operational thread-local semantics of the language describes how each thread executes.
However, the values of read operations which access shared memory locations are not concretely determined at this stage.
Instead, each operation will arbitrarily and non-deterministically pick a value to continue execution with, and generate an \textit{event} which records the choice made, and the location accessed.
Similarly, write operations to shared memory do not concretely mutate program state, but instead generate an event recording what was written, and where.

Given a complete execution of all threads at this thread-local level, the specification defines a structure called a \textit{candidate execution}.\footnote{
In C++, this is known as a \textit{pre-execution}.}
Intuitively, a candidate execution represents an execution that is consistent with
the language's sequential semantics.
It contains the set of all events generated by the thread-local semantics, together with a possible justification for the arbitrarily chosen read values, which must be checked further.

One part of this justification is a ``reads-from'' relation which must link every read event (with an arbitrarily picked value) to a write event which writes the value that was picked.
Other relations are included in the justification which enforce ordering constraints on the shape of the reads-from relation (representing, for example, inter-thread synchronization).

The language defines a second layer of axiomatic constraints over candidate executions, the \textit{axiomatic memory model}, which classifies candidate executions as either valid or invalid.
It is required by the specification that any concretely observable execution must correspond to some valid candidate execution.
In this way, the memory model determines which read/write values chosen at the level of the thread-local semantics are permitted to be observed.

%!TEX root = ../main.tex

\begin{figure*}[h]
\begin{minipage}{1\textwidth}

\begin{tabular}{rcl@{\qquad\qquad}rcl}
$\type{mode}$ & $::=$ & $\tcons{Unordered}~\CMMT~\ordun~|~\tcons{SeqCst}~\CMMT~\ordsc~|~\tcons{Init}~\CMMT~\ordi$ &
$\type{addr}$ & $::=$ & $\alpha \ldots $ an infinite set of abstract names
\end{tabular}
\end{minipage}\\[0.3\baselineskip]
\begin{minipage}{0.3\textwidth}
\begin{tabular}{rcl@{~}l@{~}l}
$\type{event}$ & $::=$ &  \{ &  \fld{ord} & : \type{mode} \\
&&& \fld{block} & : \type{addr} \\
&&& \fld{index} & : \type{nat} \\
&&& \fld{reads} & : $\listtype{\bytetype}$ \\
&&& \fld{writes} & : $\listtype{\bytetype}$ \\
&&& \fld{tearfree} & : $\booltype$ \}
\end{tabular}
\end{minipage}%
\begin{minipage}{0.7\textwidth}
\raggedleft
\begin{tabular}{rcl@{~}l@{~}l}
$\candidatetype$ & $::=$ &  \{ & \fld{evs} & : $\settype{\evttype}$ \\
&&& \text{\seqb~\CMMT~\seqbr} & : $\settype{(\evttype \times \evttype)}$ \\
&&& \multicolumn{2}{@{}l}{\asw~\CMMT~\aswr~: $\settype{(\evttype \times \evttype)}$} \\
&&& \text{\rbf~\CMMT~\rbfr} & : $\settype{(\nattype \times \evttype \times \evttype)}$ \\
&&& \text{\tot~\CMMT~\totr} & : $\settype{(\evttype \times \evttype)}$ \} \\
\\
\end{tabular}

\end{minipage}\\[0.5\baselineskip]

\begin{tabular}{rcl@{\qquad}rcl}
$\func{range}_r(\obj{E}$ : \evttype) & $\triangleq$ & [ \obj{E}.\fld{index} \ldots $\obj{E}.\fld{index} + |\obj{E}.\fld{reads}|$ ) & $\func{write}(\obj{E}$ : \evttype) & $\triangleq$ & $(\obj{E}.\fld{writes} \neq [])$ \\
$\func{range}_w(\obj{E}$ : \evttype) & $\triangleq$ & [ \obj{E}.\fld{index} \ldots $\obj{E}.\fld{index} + |\obj{E}.\fld{writes}|$ ) & $\func{overlap}(\obj{E}_1, \obj{E}_2$ : \evttype)& $\triangleq$ & $\obj{E}_1.\fld{block} = \obj{E}_2.\fld{block}~\wedge$ \\
$\func{range}(\obj{E}$ : \evttype) & $\triangleq$ & $\func{range}_r(\obj{E}) \cup \func{range}_w(\obj{E})$ &&& $\quad\func{range}(\obj{E}_1) \cap \func{range}(\obj{E}_2) \neq \emptyset$
\vspace{0.5em}
\end{tabular}\\[0\baselineskip]

\noindent{\color{lightgray} \rule{\textwidth}{0.3pt}}

\begin{minipage}[t]{0.35\textwidth}
\raggedright
\textbf{Derived relations} \textit{\small  (wrt. a candidate execution)}
\vspace{-1.4em}
$$
\begin{array}{l}
\\
  \rf~\CMMT~\rfr \triangleq \\
  \qquad\{ \xxpair{\obj{A}}{\obj{B}} \mid \exists k \ldotp \xxtriple{k}{\obj{A}}{\obj{B}} \in \rbf \}
\\[0.2em]
\hb~\CMMT~\hbr \triangleq \\[-0.2em]
\qquad\left(\pbox{\textwidth}{$ \seqb \cup \sw~\cup$ \\ $\{ ~ \xxpair{\obj{A}}{\obj{B}} \mid \obj{A}.\fld{ord} =  \tcons{Init} \wedge \func{overlap}(\obj{A}, \obj{B}) ~ \} $} \right)^{\scalebox{1.1}{$+$}}
\end{array}
$$
\end{minipage}%
\begin{minipage}[t]{0.762\textwidth}
\vspace{-1.4em}
$$
\begin{array}{l}
\\
\\
\sw~\CMMT~\swr \triangleq \\ \qquad
\begin{array}{l}
  \left\{~ \xxpair{\obj{A}}{\obj{B}} \hspace{0.5em} \left| \hspace{0.5em}
  \pbox{0.5\textwidth}{
    $\obj{A}~\rf~\obj{B} \wedge \obj{B}.\fld{ord} =  \tcons{SeqCst}~\wedge$ \\
    $\left(\pbox{0.5\textwidth}{
    $\left(\pbox{0.5\textwidth}{
      $\func{range}_w(\obj{A}) = \func{range}_r(\obj{B})~\wedge \obj{A}.\fld{ord} =  \tcons{SeqCst}$
    } \right) \vee$\\
        $\left(\pbox{0.5\textwidth}{
      $\forall \obj{C}.~\obj{C}~\rf~\obj{B} \longrightarrow \obj{C}.\fld{ord} =  \tcons{Init}$
    } \right)$
    } \right)$
  }
  \right. ~\right\}
\\
  \cup~\asw
\end{array}
\end{array}
$$
\end{minipage}
%\vspace{-0.5\baselineskip}
\caption{JavaScript Candidate Execution. We introduce short names for some relations after the ``\protect\CMMT''.}
\label{fig:candidate-execution}
\end{figure*}

\subsection{Candidate Executions}

\begin{sloppypar}
Candidate executions are formally specified in Fig.~\ref{fig:candidate-execution}.
A~candidate execution for a given thread-local execution consists of the set of events $\textbf{\fld{evs}}$, and the relations over this set $\textbf{\seqb}$, $\textbf{\asw}$, $\textbf{\rbf}$, and $\textbf{\tot}$, which form the potential justification.

Our notation mainly follows that of the formal C/C++11 memory model~\cite{Batty:2011}; we treat relations between events as sets of tuples, and make use of standard notation from relational algebra to manipulate them.
For example, the transitive closure of a binary relation \rel{rel} is given as $\rel{rel}^+$, and its inverse as $\rel{rel}^{-1}$.
For binary relations, we use an infix notation to indicate membership i.e. $\obj{A}~\rel{rel}~ \obj{B} \equiv \xxpair{\obj{A}}{\obj{B}}\in \rel{rel}$.

The candidate execution's components $\fld{evs}$, $\seqb$, and $\asw$ are all precisely determined from the thread-local execution.

Events generated by the thread-local execution are contained in $\textbf{\fld{evs}}$.
Each event records its \fld{mode}, which may be either Sequentially Consistent (atomic), unordered (non-atomic), or a specially distinguished initializing write, and the locations that it accesses: the
combination of the \fld{block}, representing the address of an
individual SharedArrayBuffer; the \fld{index}, representing the
starting position of the access within the SharedArrayBuffer; and the
\fld{reads} and \fld{writes} fields, representing the list of bytes
the event read or wrote (respectively) as determined by the
thread-local semantics. Moreover, the thread-local semantics
  marks certain events as \emph{tearfree}, which will be explained
  later.
The \fld{block} component's main function is merely to ensure that
accesses to different SharedArrayBuffers are treated as having
disjoint ranges by construction. Hence, throughout this paper we will
usually work with the assumption that all accesses in a
candidate execution are to the same block.

The $\textbf{\seqb}$ component is an intra-thread relation between events that records their order in the control-flow unfolding of the execution.
It ensures that events that occur sequentially in the same thread are strongly ordered with respect to each other.

The relation $\textbf{\asw}$ records places where the thread-local semantics' action implies strong inter-thread ordering in the memory model.
For example, when a parent thread creates a child thread, the thread-local semantics has an $\asw$ edge from the parent to the child. This edge will ensure that all previous accesses by the parent are visible to the child.

In addition, the candidate execution contains two relations which do not merely arise from the thread-local semantics.
The $\textbf{\rbf}$ component represents a possible justification for the values of read events, by relating them to write events in the execution.
It is defined in a byte-wise manner; each byte location of a multi-byte read event is related to a write event on that location, and each byte may be justified by a different write.
Therefore \mbox{$\xxtriple{k}{\obj{E}_w}{\obj{E}_r} \in \rbf$} means the event $\obj{E}_r$ reads the value of $\obj{E}_w$ at byte index $k$.

Finally, the $\textbf{\tot}$ component records some total order over all
events. Sequentially consistent atomic operations must obey certain restrictions about
where they can appear in this total order, resulting in stronger guarantees about their
behaviour.

The relations $\rbf$ and $\tot$ are arbitrarily picked when constructing the candidate execution, subject to certain intuitive well-formedness conditions ($\rbf$ must associate read events to write events with the same byte values, $\tot$ must be a strict total order on events) which we define explicitly in a supplementary appendix~\cite{papersuppl}.
At the programmer level, an execution is only observable if, for some choice of $\rbf$ and $\tot$, a candidate execution exists which is allowed by the memory model.

Fig.~\ref{fig:candidate-execution} also defines three derived
relations: intuitively, the $\rf$ relation recovers a C/C++11-style
event-to-event definition from the $\rbf$ relation, by projecting
away the byte
index component%
%% \footnote{Note that, by convention, and in common with C/C++11, we make the write the \textit{left} component of the \rf relation. This makes the model neater, at the cost that conditions involving \rf become less intuitive to pronounce.};
\footnote{
By convention, and in common with C/C++11, we make the write the \emph{left} component of the \rf relation.};
 %% This makes the model neater, at the cost of diverging from the JavaScript specification's presentation that conditions involving \rf become less intuitive to pronounce.};
%
the $\sw$ relation records the extra synchronization
guarantees made by \tcons{SeqCst} atomics; the $\hb$ relation is the
transitive closure of different ordering constraints.

Throughout the paper, we will give graphs representing (fragments of) candidate executions.
We give a simple example in Fig.~\ref{fig:MP-simple-exec}, which depicts a valid candidate execution, including relevant derived relations, for Fig,~\ref{fig:MP-simple}, which justifies the outcome \mbox{\texttt{r0 = 5 $\wedge$ r1 = 3}}.
Each event in the candidate execution corresponds to a load/store operation performed by the thread-local semantics.
Some relation edges are elided where irrelevant (for example, the precise choice of $\totr$ is not interesting in this example) or otherwise obvious ($\seqbr$ is trivial from the program layout).
\end{sloppypar}

\subsection{Valid Candidate Executions}

%!TEX root = ../main.tex

\begin{figure*}[h]
\raggedright
\figLeftMargin
% Having the text here makes it unclear whether it's part of the
% normal flow of text or the figure. Also "iff" is not true: still
% subject to well-formedness
%
% A candidate execution is considered \textit{valid} iff it satisfies all the following properties.
% \\[1em]
\begin{minipage}{0.479\textwidth}
\textbf{Happens-Before Consistency (1):}\\
$\hb \subseteq \tot$
\end{minipage}%
\begin{minipage}{0.521\textwidth}
\textbf{Happens-Before Consistency (2):}\\
$\forall \obj{E}_w \obj{E}_r \ldotp \obj{E}_w~\rf~\obj{E}_r \longrightarrow \neg (\obj{E}_r~\hb~\obj{E}_w)$
\end{minipage}
\\[1em]
\begin{minipage}{0.479\textwidth}
\begin{tabular}{@{}l}
\textbf{Happens-Before Consistency (3):}\\
$\forall \xxtriple{k}{\obj{E}_w}{\obj{E}_r} \in \rbf \ldotp$\\
\qquad$\nexists \obj{E}'_w \ldotp (\obj{E}_w~\hb~\obj{E}'_w)~\wedge$\\
\qquad\qquad\quad$(\obj{E}'_w~\hb~\obj{E}_r) \wedge k \in \func{range}_w(\obj{E}'_w)$
\end{tabular}
\end{minipage}%
\begin{minipage}{0.521\textwidth}
\begin{tabular}{@{}l}
\textbf{Tear-Free Reads:}\\
$\forall \obj{E}_r \ldotp \obj{E}_r.\fld{tearfree}~\longrightarrow$\\
\qquad$\left|\left\{~
\obj{E}_w~\left|~
\pbox{0.5\textwidth}{$\obj{E}_w~\rf~\obj{E}_r~\wedge~\obj{E}_w.\fld{tearfree}~\wedge$ \\ $\func{range}_w(\obj{E}_w) = \func{range}_r(\obj{E}_r)$}
\right.~
\right\}\right| \leq 1$
\end{tabular}
\end{minipage}
\\[1em]
\textbf{Sequentially Consistent Atomics (first attempt):}\\
$\forall \obj{E}_w \obj{E}_r \ldotp \obj{E}_w~\sw~\obj{E}_r \longrightarrow %$\\
%$
\quad \nexists \obj{E}'_w \ldotp (\obj{E}_w~\tot~\obj{E}'_w) \wedge  (\obj{E}'_w~\tot~\obj{E}_r) \wedge \func{range}_w(\obj{E}'_w) = \func{range}_r(\obj{E}_r)$

\caption{Candidate execution validity as defined by the latest JavaScript specification.}
\label{fig:candidate-execution-valid}
\end{figure*}

As discussed, an execution is allowed by the specification if it is possible to pick $\rbf$ and $\tot$ relations (in C/C++11 called an \textit{execution witness}~\cite{Batty:2011}) such that the resulting candidate execution is valid.
Validity of a candidate execution is defined in
Fig.~\ref{fig:candidate-execution-valid}.

\paragraph{Happens-Before Consistency (1-3)}%
  An edge in \hb implies a strong ordering constraint in the model. Rule (1) states that the total order $\totr$ must contain $\hbr$.
  The other rules ensure that reads do not, under any circumstances, observe
  writes in a way that is inconsistent with \hb; (2) a read cannot be \hb a write it reads from; (3) 
  a read $E_r$ cannot read ``stale'' bytes from a write $E_w$ if there is a ``newer'' write $E_w'$ according to \hb.

\paragraph{Tear-Free Reads}%
  This rule provides extra guarantees on the behaviour of events 
  marked as \fld{tearfree}.
  A \emph{tearing} event (one that is not \fld{tearfree}) represents
  an access which be may be observed as a series of smaller
  independent accesses.
  One example would be a 64-bit access implemented on a 32-bit machine
  as a pair of 32-bit accesses.

  For events declared as \fld{tearfree}, this rule guarantees that
  \fld{tearfree} reads will never read from more than one
  \fld{tearfree} write of the same size and alignment.
  That is, it will not observe an interleaving of bytes from multiple
  \fld{tearfree} writes.
\paragraph{Sequentially Consistent Atomics} %
  Finally, the SC Atomics rule is intended to further restrict
  \tcons{SeqCst} atomics, so that \tcons{SeqCst} reads observing
  \tcons{SeqCst} writes obey the \tot.
  Note that it is still possible for more relaxed behaviour to occur
  if an \tcons{Unordered} access is intermingled with \tcons{SeqCst}
  ones.
  
As we will show in~\S\ref{sec:deficiencies}, this last condition
must be re-written to correct deficiencies in the model.

JavaScript's specification describes the memory model in a precise, semi-formal pseudocode.
When rendering the model in logic, it is convenient for us to make some changes in presentation
that do not affect the model.
These are discussed in a supplemental appendix~\cite{papersuppl}.

\section{Corrected Model Deficiencies}
\label{sec:deficiencies}

The current JavaScript concurrency model contains two major
deficiencies that will be discussed in the following, along with our
proposed alterations to the model.
These proposals have now been accepted by the JavaScript committee for inclusion in an upcoming edition of the standard.

Throughout this paper, we will work with a restricted fragment of the JavaScript language, consisting only of programs with a fixed number of threads, in which each thread has only shared memory accesses and simple control-flow, and where the program contains an already-initialised SharedArrayBuffer (potentially wrapped by multiple typed arrays). We assume that the initialisation is done before all other accesses, since the concurrent behaviour of memory allocation relies on the (relaxed) behaviour of dynamic allocation involving OS calls, which is beyond the scope of this work to reason about.
This fragment is sufficient to exhibit all the deficiencies discussed in the next sections, and later, verify their absence in the revised model incorporating our fixes.
This language fragment is the JavaScript equivalent to the fragment of C/C++11 considered by previous work such as~\cite{Podkopaev:2019:BGP:3302515.3290382} and \cite{Lahav:2017:RSC:3062341.3062352}, with the additional complication that our accesses may be mixed-size.

After discussing the details of the deficiencies of the current
JavaScript model, we show our use of Alloy for generating
counter-examples for the original model in~\S\ref{sec:alloy}, and we
detail our Coq verification of the corrected model
in~\S\ref{sec:coq}.  We also discover
problems with another feature of the language, relating to thread
suspension, which we describe in \S\ref{sec:wait-notify}.

\subsection{ARMv8 Compilation}
\label{sec:arm_compilation_failure}

\renewcommand\thefigure{6}
%!TEX root = ../main.tex
%
\begin{figure*}[h]
\begin{subfigure}{\textwidth}
\begin{tabular}{p{0.55\textwidth}p{0.4\textwidth}}
  \centering
  \begin{tikzpicture}[%
    /litmus/append to tikz path search,
    kind=assem,
    JavaScript,
    threads baseline,
  ]
  % Thread 0
  \node[simple instructions] {
    |Atomics.store(b,0,1);  | \\
    |r1 = Atomics.load(b,1);| \\
    |// r1 = 1              | \\
  };

  % Thread 1
  \node[simple instructions] {
    |Atomics.store(b,1,1);  | \\
    |Atomics.store(b,1,2);  | \\
    |b[0] = 2;              | \\
    |r2 = Atomics.load(b,0);| \\
    |// r2 = 1              | \\
  };
  \end{tikzpicture}
  & \hspace{-1cm}
  \begin{tikzpicture}[%
    /litmus/append to tikz path search,
    kind=events,
    JavaScript,
    threads baseline,
  ]
  % Thread 0
  \node[simple instructions] {
    a:W\textsubscript{SC} b[0..3]=1 \\
    b:R\textsubscript{SC} b[4..7]=1 \\
  };

  % Thread 1
  \node[simple instructions] {
    c:W\textsubscript{SC} b[4..7]=1 \\
    d:W\textsubscript{SC} b[4..7]=2 \\
    e:W\textsubscript{Un} b[0..3]=2 \\
    f:R\textsubscript{SC} b[0..3]=1 \\
  };

  % [rf, out=-165, in=175, looseness=1.45, swap, label pos=0.75]
  % disable edge routing
  \draw (a) edge[tot, out=-170, in=180, looseness=1.5, label pos=0.75, swap] (e)
        (b) edge[tot, swap] (d)
        (c) edge[rf, swap] (b)
        (e) edge[disable edge routing, tot, out=0, in=0, looseness=2] (f)
        (f) edge[multi'={sw^{-1},rf^{-1}}, color=sw color, out=180, in=-175, looseness=1.75, label pos=0.2] (a);
        
  \path[use as bounding box] (-2.15,-1.4) rectangle (5.35,0); % adjust to fit
     
east);
  \end{tikzpicture}
  \end{tabular}
\caption{An outcome forbidden by JavaScript. Note the
  shape of Fig.~\ref{fig:sc-shape1} appearing between events a, e, and
  f.}
\label{fig:armv8-counter-a}
\vspace{-0.3\baselineskip}
\end{subfigure}

\begin{subfigure}{\textwidth}
  \bigskip
  \begin{tabular}{p{0.55\textwidth}p{0.4\textwidth}}
  \centering
  \begin{tikzpicture}[%
    /litmus/append to tikz path search,
    kind=assem,
    AArch64,
    threads baseline,
  ]
  % Thread 0
  \node[simple instructions] {
    |stlr W0,[X1]| \\
    |ldar W2,[X3]| \\
    |// W2 = 1   | \\
  };

  % Thread 1
  \node[simple instructions] {
    |stlr W0,[X3]| \\
    |stlr W2,[X3]| \\
    |str  W2,[X1]| \\
    |ldar W4,[X1]| \\
    |// W4 = 1   | \\
  };
  \end{tikzpicture}
  & \hspace{-1cm}
  \begin{tikzpicture}[%
    /litmus/append to tikz path search,
    kind=events,
    AArch64,
    threads baseline,
  ]
  % Thread 0
  \node[simple instructions] {
    a:W\textsubscript{rel} b[0..3]=1 \\
    b:R\textsubscript{acq} b[4..7]=1 \\
  };

  % Thread 1
  \node[simple instructions] {
    c:W\textsubscript{rel} b[4..7]=1 \\
    d:W\textsubscript{rel} b[4..7]=2 \\
    e:W b[0..3]=2 \\
    f:R\textsubscript{acq} b[0..3]=1 \\
  };

  \draw (a) edge[rf, out=-175, in=180, looseness=1.75, swap, label pos=0.8] (f)
        (b) edge[fr, swap] (d)
        (c) edge[rf, swap] (b)
        (c) edge[disable edge routing, co, out=0, in=0, looseness=2] (d)
        (e) edge[disable edge routing, co, out=180, in=-170, looseness=1.5, swap, label pos=0.2] (a);
        
  \path[use as bounding box] (-2.1,0) rectangle (5.25,0); % adjust to fit
  \end{tikzpicture}
\end{tabular}
\caption{When the program is compiled to ARMv8, the outcome is allowed. }
\label{fig:armv8-counter-b}
\end{subfigure}
\vspace{-0.5\baselineskip}
\caption{A JavaScript program which violates the memory model when compiled to ARMv8.}
\label{fig:armv8-counter}
\vspace{-0.2\baselineskip}
\end{figure*}

The ARMv8 architecture provides the Load Acquire (\texttt{ldar}) and Store
Release (\texttt{stlr}) instructions, memory access instructions with
certain thread-local ordering guarantees, which are also intended as compilation targets for
C/C++ sequentially consistent atomics (\texttt{memory\_order\_seq\_cst}).
It was intended that the JavaScript model should support this compilation scheme, which is implemented in at least one Web browser (Chrome). We have identified, however, that the current JavaScript memory model, as presented in Fig.~\ref{fig:candidate-execution-valid}, is incompatible with this compilation scheme.
The key issue is the \textbf{Sequentially Consistent Atomics} condition, which, in addition to restricting \ordsc accesses, also restricts \ordun accesses by disallowing executions of the shape shown below.
% in Fig.~\ref{fig:sc-shape1}, where the three events all access the same byte range.

% TODO: figures 5 to 10 numbered manually
\renewcommand\thefigure{5}
\begin{figure}[H]
\begin{tikzpicture}[%
  /litmus/append to tikz path search,
  kind=events,
  JavaScript,
  show event label=false,
  /litmus/every relation/.style=,
]
  \node[simple instructions without header] {
    a:W\textsubscript{SC} b[$i..j$] \\
  };
  \node[simple instructions without header] {
    b:W\textsubscript{Un} b[$i..j$] \\
  };
  \node[simple instructions without header] {
    c:R\textsubscript{SC} b[$i..j$] \\
  };

  \draw (a) edge[tot] (b)
        (b) edge[tot] (c)
        (c) edge[sw^{-1}, out=-168, in=-12] (a);
\end{tikzpicture}
\vspace{-0.5\baselineskip}
\caption{Forbidden by \textbf{SC Atomics (first attempt)}.}
\label{fig:sc-shape1}
\end{figure}

It is possible to craft a program that produces a particular output
only in a candidate execution that contains this shape forbidden by
JavaScript. Nevertheless, the behaviour is observable on ARMv8 when
using the compilation scheme that maps \ordun accesses to bare ARMv8
accesses and \ordsc accesses to ARMv8 release/acquire accesses.
Such a program is shown in Fig.~\ref{fig:armv8-counter}.
The candidate execution of Fig.~\ref{fig:armv8-counter-a} is (in the unfixed JavaScript model) forbidden because it includes the shape of Fig.~\ref{fig:sc-shape1}.
Note that no other candidate execution can make this output observable, since alternative configurations of edges are also forbidden by the memory model.

In particular, because the event (b) reads 1, there must be a \totr~edge from (b) to the write (d).
If the edge were the other way around, (b) would not be allowed to read 1, and could only read 2 from (d), since reading from (c) would be forbidden by the \textbf{Sequentially Consistent Atomics} rule.
Therefore the \totr~edge from (a) to (e) is also fixed, because of \totr's transitivity and the fact that $\hbr \in \totr$.

We originally discovered a larger counter-example by hand; this small counter-example was found automatically as part of our Alloy model-checking efforts, as detailed in~\S\ref{sec:alloy}.
%
%\todocp{Instead: ``We discovered this program when investigating the JavaScript compilation scheme to ARMv8 (manually) and our Alloy model-checking search is able to automatically re-find it, as detailed in~\S\ref{sec:alloy}.''}
% Conrad: the hand-found example was much bigger - this one was found by Alloy
%
We have confirmed that the corresponding execution is architecturally
allowed in ARMv8 for the compiled program, by running it in the two
existing executable concurrency models for ARMv8
\cite{deacon-cat,PulteEtAl2018}. We are also able to
observe this execution experimentally, with the caveat that we must use WebAssembly to force efficient compilation (see \S\ref{sec:experimental}).

\paragraph{Proposed Fix}
We propose a weakening of the JavaScript model which permits this ARMv8 compilation scheme:
weakening the \textbf{Sequentially Consistent Atomics} condition of Fig.~\ref{fig:candidate-execution-valid} as follows:
\begin{figure}[H]
\vspace{-0.5\baselineskip}
\raggedright
\textbf{SC Atomics (second attempt):}\\
$\forall \obj{E}_w \obj{E}_r \ldotp \obj{E}_w~\sw~\obj{E}_r \longrightarrow$\\
$\quad\nexists \obj{E}'_w \ldotp \obj{E}'_w.\fld{ord} =  \tcons{SeqCst} \wedge (\obj{E}_w~\tot~\obj{E}'_w)~\wedge$\\
$\qquad\qquad(\obj{E}'_w~\tot~\obj{E}_r) \wedge \func{range}_w(\obj{E}'_w) = \func{range}_r(\obj{E}_r)$
%\vspace{-0.5\baselineskip}
\end{figure}

This means that shapes like Fig.~\ref{fig:sc-shape1} are no longer forbidden.
Intuitively, the original condition was putting an ordering constraint on \ordun accesses that were part of a data-race (see~\S\ref{sec:scdrf}), by forcing them to have a certain position in \totr, even when that position was not enforced by \hbr.
The ARMv8 \texttt{ldar/stlr} instructions were designed to support C/C++11 atomics where such a data-race would be undefined behaviour, and they do not provide the guarantees necessary for (uncorrected) JavaScript's \ordun ordering in the racy case.

\subsection{SC-DRF}
\label{sec:scdrf}

\citet{weakeningwasm} identified that the JavaScript model does not
provide Sequential Consistency (SC) of Data-Race-Free programs (SC-DRF), an
important correctness condition of the relaxed memory model that
JavaScript intends to provide. After a discussion of JavaScript's
choice of SC-DRF definition, we detail JavaScript's violation of this
SC-DRF property and integrate their proposed correction with our ARMv8 fix,
for subsequent verification.
% make it clearer what is Watt [44] and what we do

Informally, the SC-DRF property says that a data-race-free program
will only give rise to SC results, i.e. results corresponding to a
sequential interleaving of its accesses~\cite{Lamport:1979:MMC:1311099.1311750}. This is an important property
because it allows programmers to reason about their software under a
simpler semantics: so long as they ensure their programs are
data-race-free, they can program according to the simpler SC model.

\renewcommand\thefigure{10}
\begin{figure*}[b]
%\vspace{0.2\baselineskip}
\footnoterule
\vspace{0.5\baselineskip}
\begin{minipage}{\textwidth}
\raggedright
\figLeftMargin
\textbf{Sequentially Consistent Atomics (final):}\\
$\forall \obj{E}_w \obj{E}_r \ldotp
\obj{E}_w~\rf~\obj{E}_r \;\wedge\;
\obj{E}_w~\hb~\obj{E}_r
\longrightarrow$\\
$\quad \nexists \obj{E}'_w \ldotp \obj{E}'_w.\fld{ord} =  \tcons{SeqCst} \;\wedge\; \obj{E}_w~\tot~\obj{E}'_w
\;\wedge\; \obj{E}'_w~\tot~\obj{E}_r
~ \wedge$\\
$\qquad \qquad \left(
\begin{array}{r@{\,}l@{\,}l}
 & (\func{range}_w(\obj{E}'_w) = \func{range}_r(\obj{E}_r)
         &\wedge\; \obj{E}_w~\sw~\obj{E}_r )\\
\vee&
  (\func{range}_w(\obj{E}_w) = \func{range}_w(\obj{E}'_w)
         &\wedge\; \obj{E}_w.\fld{ord} = \tcons{SeqCst}
         \;\wedge\; \obj{E}'_w~\hb~\obj{E}_r )\\
\vee&
  (\func{range}_w(\obj{E}'_w) = \func{range}_r(\obj{E}_r)
         &\wedge\; \obj{E}_w~\hb~\obj{E}'_w
         \;\wedge\; ~\obj{E}_r.\fld{ord} = \tcons{SeqCst})
\end{array}\right)$
\end{minipage}
%\vspace{-0.5\baselineskip}
\caption{The \textbf{Sequentially Consistent Atomics} rule containing all proposed fixes.}
\label{fig:sc-atomics-final}
\vspace{-0.2\baselineskip}
\end{figure*}

\paragraph{Discussion of SC-DRF definition}
The JavaScript standard explicitly specifies the SC-DRF property it intends to provide.
Their specification of SC-DRF is analogous to the statement of the property given in the C++11 standard, and used by \citet{Batty:2012}:
informally, two JavaScript accesses are considered to data-race if they overlap, at least one of them is a write, they are not both same-range \ordsc atomics, and the two accesses are not ordered by \hbr. The formal definition is given in Fig.~\ref{fig:data-race}.
%
%\vspace{-0.5em}
\renewcommand\thefigure{7}
%!TEX root = ../main.tex
%
\begin{figure}[H]
\raggedright
\textbf{Data-Race:} (for two events \obj{A} and \obj{B} in a given CE)\\
$\quad(\obj{A}.\fld{ord} = \ordun \vee \obj{B}.\fld{ord} = \ordun \vee \func{range}(\obj{A}) \neq \func{range}(\obj{B}))$ $\wedge$ \\
$\qquad\func{overlap}(\obj{A}, \obj{B}) \wedge (\func{write}(\obj{A}) \vee \func{write}(\obj{B}))~\wedge$\\
$\quad\qquad\neg(\obj{A}~\hb~\obj{B} \vee \obj{B}~\hb~\obj{A})$
%
%\vspace{-0.5\baselineskip}
\caption{Definition of a JavaScript data-race.}
\label{fig:data-race}
%\vspace{-0.2\baselineskip}
\end{figure}%
\vspace{-0.5em}
A program is then called data-race-free if it has no (JavaScript-allowed) execution
containing a data-race, and the JavaScript specification says that such a
data-race-free program should only have SC behaviours.

A ``model-agnostic'' definition of SC-DRF has since been proposed (but not adopted) for C/C++~\cite{10.1007/978-3-662-46669-8_12}, based on a simpler definition of data-race-freedom that requires the absence of data-races only in \emph{Sequentially Consistent executions}, instead of every possible execution allowed by the model.
This paper concentrates exclusively on JavaScript's (and by extension, C++11's) formulation, which we refer to as ``model-internal SC-DRF'' where appropriate, in order to disambiguate our verification claims.

%Note that C++ and JavaScript differ on their guarantees if a program \textit{does} contain a data-race.
%%
%In C++, a program with a non-atomic data-race has entirely undefined behaviour; JavaScript's model gives such a program defined behaviour, but this defined behaviour may be extremely weak.

\renewcommand\thefigure{8}
%!TEX root = ../main.tex

\begin{figure}[H]
\begin{tikzpicture}[%
  /litmus/append to tikz path search,
  kind=assem,
  JavaScript,
  threads baseline,
]

% Thread 0
\node[simple instructions] {
  |A-.store(b, 0, 1);| \\
};

% Thread 1
\node[simple instructions] {
  |A-.store(b, 0, 2);       | \\
  |if (A-.load(b, 0) == 1) {| \\
  |  r = b[0]; //r=2             | \\
  |}                             | \\
};
\end{tikzpicture}\\[-0.5\baselineskip]
%\hspace{-1.1cm}
\begin{tikzpicture}[%
  /litmus/append to tikz path search,
  kind=events,
  JavaScript,
  threads baseline,
]

% Thread 0
\node[simple instructions] {
  a:W\textsubscript{SC} b[0..3]=1 \\
};

% Thread 1
\node[simple instructions] {
  b:W\textsubscript{SC} b[0..3]=2 \\
  c:R\textsubscript{SC} b[0..3]=1 \\
  d:R\textsubscript{Un} b[0..3]=2 \\
};

\draw (a) edge[sw, out=-50, in=180, looseness=0.5] (c)
      (a) edge[hb, out=-60, in=180] (d)
      (b) edge[tot, swap] (a)
      (d.east) edge[multi'={hb^{-1},rf^{-1}}, out=0, in=0, looseness=2.2, swap, label pos=0.5] (b.east);
      
     \path[use as bounding box] (-1.7,0) rectangle (6.3,0); % adjust to fit
\end{tikzpicture}
\vspace{-0.2\baselineskip}
\caption{SC-DRF violation by JavaScript program.}
\label{fig:sc-drf-example}
\vspace{-0.2\baselineskip}
\end{figure}

\paragraph{JavaScript SC-DRF failure}
The JavaScript specification claims that the model is SC-DRF. However,
as described by \citet{weakeningwasm}, it is possible to give a
counter-example: a program that is data-race-free, but nevertheless has
an execution which cannot be explained as a sequential interleaving of the program's accesses.

That paper describes a 6 event, 2 (distinct) location counter-example.
Using the Alloy search of~\S\ref{sec:alloy}, we are able to
find a 4 event, 1 location counter-example (Fig.~\ref{fig:sc-drf-example}).
No sequential interleaving of the program's accesses can explain why the non-atomic load of thread 1 can read 2.

\pagebreak

\citet{weakeningwasm} propose a strengthening of the model that would restore SC-DRF, by adding two sub-conditions to \textbf{Sequentially Consistent Atomics}.
These disallow the two shapes shown in Fig.~\ref{fig:sc-shape2}.

\renewcommand\thefigure{9}
\begin{figure}[h]
\begin{minipage}{0.404\textwidth}
\begin{tikzpicture}[%
  /litmus/append to tikz path search,
  kind=events,
  JavaScript,
  show event label=false,
  threads distance=8mm,
  /litmus/every relation/.style=,
]
  \node[simple instructions without header] {
    a:W\textsubscript{SC} b[$i..j$] \\
  };
  \node[simple instructions without header] {
    b:W\textsubscript{SC} b[$i..j$] \\
  };
  \node[simple instructions without header] {
    c:R\textsubscript{any} b[$i'..j'$] \\
  };

  \draw (a) edge[tot] (b)
        (b) edge[hb] (c)
        (c) edge[multi'={rf^{-1},hb^{-1}}, out=-168, in=-12] (a);
\end{tikzpicture}
\end{minipage}\\
\begin{minipage}{0.4\textwidth}
\begin{tikzpicture}[%
  /litmus/append to tikz path search,
  kind=events,
  JavaScript,
  show event label=false,
  threads distance=8mm,
  /litmus/every relation/.style=,
]
  \node[simple instructions without header] {
    a:W\textsubscript{any} b[$i'..j'$] \\
  };
  \node[simple instructions without header] {
    b:W\textsubscript{SC} b[$i..j$] \\
  };
  \node[simple instructions without header] {
    c:R\textsubscript{SC} b[$i..j$] \\
  };

  \draw (a) edge[hb] (b)
        (b) edge[tot] (c)
        (c) edge[multi'={rf^{-1},hb^{-1}}, out=-168, in=-12] (a);
\end{tikzpicture}
\end{minipage}
\caption{SC-DRF violations forbidden by the revised rule.}
\label{fig:sc-shape2}
\end{figure}
\renewcommand{\thefigure}{\arabic{figure}}
\setcounter{figure}{10}

Combining this proposal with our ARMv8 fix, we arrive at the version of the \textbf{Sequentially Consistent Atomics} condition that we proposed to the standards committee, and which has been adopted for future inclusion (Fig.~\ref{fig:sc-atomics-final}).
This new condition is neither stronger nor weaker than the original formulation.
The ARMv8 fix weakens the model, allows some previously forbidden executions.
The SC-DRF fix strengthens the model, forbidding some previously allowed executions.
We verify in Coq that the revised model supports the desired ARMv8 compilation scheme, and provides model-internal SC-DRF (\S\ref{sec:coq}).

Moreover, we identify that this condition allows another part of the model to be simplified.
The model's definition of \sw includes a special case for \KK{Init} events, ensuring that the below shape is forbidden by the \textbf{Sequentially Consistent Atomics} rule.
%to forbid the shape in Fig.~\ref{fig:sc-shape3} with the \textbf{Sequentially Consistent Atomics} rule.
%
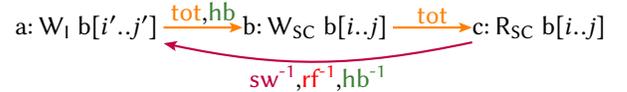
\begin{figure}[H]
\begin{tikzpicture}[%
  /litmus/append to tikz path search,
  kind=events,
  JavaScript,
  /litmus/every relation/.style=,
]
  \node[simple instructions without header] {
    a:W\textsubscript{I} b[$i'..j'$] \\
  };
  \node[simple instructions without header] {
    b:W\textsubscript{SC} b[$i..j$] \\
  };
  \node[simple instructions without header] {
    c:R\textsubscript{SC} b[$i..j$] \\
  };

  \draw (a) edge[multi'={tot,hb},color=tot color] (b)
        (b) edge[tot] (c)
        (c) edge[multi'={sw^{-1},rf^{-1},hb^{-1}}, color=sw color, out=-168, in=-12] (a);
\end{tikzpicture}
%\caption{A shape forbidden as a result of \ordi events being included in \sw}
%\label{fig:sc-shape3}
\end{figure}

Note that it is always guaranteed that $a~\hbr~b$ and $a~\hbr~c$ by the definition of \hb. Also, $\swr \subseteq \rfr$.
Therefore, the revised definition of \textbf{Sequentially Consistent Atomics} already forbids a more general shape (the second shape of Fig.~\ref{fig:sc-shape2}) and we can remove this special case, simplifying the definition of \sw, as shown below.
\begin{figure}[H]
\raggedright
$\sw~\CMMT~\swr \triangleq$ \\
$\quad\pbox{0.5\textwidth}{
  $\left\{~ \xxpair{\obj{A}}{\obj{B}} \hspace{0.25em} \left| \hspace{0.5em}
  \pbox{0.5\textwidth}{
    $\obj{A}~\rf~\obj{B} \wedge \func{range}_w(\obj{A}) = \func{range}_r(\obj{B})~\wedge$ \\
    $\obj{A}.\fld{ord}= \obj{B}.\fld{ord} =  \tcons{SeqCst}$
  }
  \right. ~\right\}$ \\
  $\cup~\asw$
}$
%\vspace{1\baselineskip}
\end{figure}

\subsection{Experimental Observations}
\label{sec:experimental}

As discussed, the ARMv8 model specifies that the execution of
Fig.~\ref{fig:armv8-counter-a} is architecturally allowed,
and so potentially observable when the code is run in the V8
JavaScript engine, a component of the Chrome Web browser, that uses
the release/acquire ARMv8 compilation scheme.
%
% tried to adapt it to mostly match that. The "is observable" part is
% completely true like that, we didn't have any observation of this
% test until we ran it on the website
%
% \todojp{As discussed, the ARMv8 model specifies that the execution of Fig.~\ref{fig:armv8-counter-a} is allowed in the ARMv8 architecture, and it is observable on hardware, so it is natural to ask whether it is also observable when the code is run in the V8 JavaScript engine, a component of the Chrome Web browser, that uses the release/acquire ARMv8 compilation scheme}
%
% We attempted to observe this behaviour end-to-end, as a JavaScript fragment running in a website, but were unsuccessful.
We attempted to observe this behaviour ``end-to-end'', by building a website running the JavaScript fragment, but were unsuccessful.
JavaScript compilation is complex, and incorporates profile-guided optimisation.
We found that we could not coax the engine to generate the efficient ARMv8 code of Fig.~\ref{fig:armv8-counter}.
However, we can take advantage of the fact that WebAssembly's memory model (for this language fragment) is designed to be identical to JavaScript's~\cite{weakeningwasm}; the exact same accesses are available as WebAssembly instructions.
Indeed, V8 compiles JavaScript and WebAssembly through the same backend.

The predictability of WebAssembly compilation as a proxy for perfectly optimised JavaScript was previously taken advantage of by the RIDL MDS attacks~\cite{ridl}. %
% \todocp{is it important to cite this here?}\todojp{I put it there because (1) it makes it clear we haven't invented the idea, and (2) it's kind of funny to use it ``for good''}
%
% CP: Ok
%
Here, instead, we use it to gain more predictability over the compilation of the litmus test.
By embedding the same test, in WebAssembly, on a website, we were able
to observe the problematic ARMv8 behaviour in Chrome, %through a web browser,
on
the LG G Flex2 H955 phone, an Android phone with a Qualcomm
Snapdragon810 SoC (quad core ARM Cortex-A57 + quad core ARM
Cortex-A53).
Due to the general shape of the test, we conjecture that any CPU exhibiting the \texttt{R+polp+pola} litmus test~\cite{litmustable} should also exhibit the counter-example behaviour.
%
%which is the model that \cite{Flur:2017:MCA:3009837.3009839} had observed it directly on.

% "even" makes it sound like this isn't nice already
%
Our experimental evidence was sufficient to motivate to the JavaScript
committee that this was a
%% actual, 
practical problem that needed to be
addressed, as they aim for the JavaScript and WebAssembly
accesses to have identical semantics.
%
% \todocp{I also like the idea of doing the crowdsourcing, but I think we should probably drop the following sentence, it's a bit distracting. }
% \todojp{It is currently repeated in the future work section, so we can drop it here?}
% We believe that this approach could be taken further to crowdsource litmus testing,
% thereby relieving researchers from the need to acquire many expensive devices to test them.
%
% CP: Dropping this for now then.

Before detailing our Alloy-based counter-example search and
model-checking of the ARMv8 compilation scheme and the SC-DRF
property for the fixed model in \S\ref{sec:alloy}, we now
discuss our work on defining a mixed-size ARMv8 model.

%!TEX root = main.tex
\section{ARMv8 Mixed-Size Model}
\label{arm_mixed}

In order to enable the Alloy-based counter-example search and bounded
verification of \S\ref{sec:alloy}, and the Coq compilation scheme correctness proofs of \S\ref{sec:coq}, we define and validate a
mixed-size ARMv8 axiomatic model, as an extension of the existing
ARMv8 axiomatic reference model.

The two starting points for developing the mixed-size axiomatic model
are the existing Flat model \cite{PulteEtAl2018}, an %% executable
operational model with mixed-size support, and ARM's reference model
\cite{deacon-cat,PulteEtAl2018},
an %% executable
axiomatic specification defined in herd \cite{Alglave:2014}, without mixed-size
support. The two models are based on extensive past research on
architectural %% relaxed-memory
concurrency for ARM (and related Power),
discussion with architects, and experimental hardware testing
\cite{PulteEtAl2018, Flur:2017:MCA:3009837.3009839,
  DBLP:conf/popl/FlurGPSSMDS16, GrayKMPSS2015, Alglave:2014,
  MarangetSS2012, SarkarEtAl2012, MadorHaimMSMAOAMS2012,
  AlglaveMSS2011, SarkarEtAl2011, AlglaveFIMSSZN2009,
  ChongIshtiaq2008, AdirAS2003, CorellaSB1993, diy7}.   The mixed-size
axiomatic model we arrive at is a generalisation of the reference
axiomatic model to mixed-size programs in a way that aims to follow
the Flat model's behaviour --- Flat has been developed in
collaboration with ARM and is extensively experimentally validated.

Our goal in developing the axiomatic mixed-size ARMv8 model is primarily to investigate JavaScript's compilation correctness.
In cases where Flat's mixed-size semantics is still potentially subject to change we choose weaker behaviours, and
it is possible that our model allows some mixed-size behaviours which are not allowed by Flat.
As long as our model is \textit{no stronger than} Flat, however,
any compilation scheme our ARMv8 model supports will also be supported by the Flat model.
We extensively validate this property experimentally, on a large corpus of tests.

In \citet{PulteEtAl2018}, the uni-size axiomatic and Flat operational
model were hand-proved equivalent (for uni-size input
programs). 
Formally proving a correspondence between mixed-size Flat and a
mixed-size axiomatic model would be a substantial effort in its own
right: extending the axiomatic model to mixed-size accesses breaks
some assumptions made by the existing proof. Extending the proof is
beyond the scope of this paper where our focus is JavaScript, and further work still needs to be done in order to find axiomatic rules that
are precisely equivalent to Flat.
However, we believe that our approach of generalising an existing uni-size axiomatic model,
combined with extensive validation, represents an important first step in solving this more general problem.

\subsection{Validation}

The experimental validation is based on the corpus of 11,587 existing litmus
tests from prior work on ARMv8 (the majority systematically generated
with \texttt{diy}~\cite{diy7}, and including hand-written tests used
in
\citet{Flur:2017:MCA:3009837.3009839,DBLP:conf/popl/FlurGPSSMDS16}).
We run the Flat model on this test suite and enumerate, for each
test, the set of all behaviours allowed by Flat. We
instrumented the Flat model to generate, for each such possible
outcome, the candidate execution corresponding to the operational
model's trace. We log the candidate executions, and feed them
into the Alloy-based ARMv8 axiomatic model to ensure the
\emph{soundness} of the axiomatic model: that it allows each such Flat-allowed
execution.

The litmus test suite we run contains 11,587 litmus tests. We run the
tests on a Ubuntu 18.04.2 POWER9 machine (160 CPUs at 2.9GHz, 125GB
ram) with no memory limit and 168h time limit. Of the
11,587 tests, 11,578 complete in Flat (2635 mixed-size
and 8943 non-mixed-size), so all but 9. Of these 9, 3 are
due to instructions currently unsupported by Flat, 4 running
out of memory, 1 running out of time, and another test crashing with
an unspecified error.
For the 11,578 tests where Flat successfully
completes, it generates a total of 167,014 candidate executions. We
run the mixed-size Alloy-based ARMv8 axiomatic model on these and confirm that
it allows every such Flat-allowed execution.

\section{Alloy Verification}
\label{sec:alloy}

For the SC-DRF and ARMv8 compilation issues described in \S\ref{sec:scdrf} and \S\ref{sec:arm_compilation_failure}, we define the JavaScript and mixed-size ARMv8 models in the Alloy model checker~\cite{Jackson:2002}, allowing us to compare the two models and investigate whether individual litmus tests are allowed by the models.
This approach was first used
by~\citet{Wickerson:2017:ACM:3009837.3009838}, who took existing
uni-size models, written in herd~\cite{Alglave:2014}, and automatically converted them to Alloy.
In contrast, we directly transcribe the JavaScript (corrected and uncorrected) and ARMv8 models into Alloy by hand.
Alloy's syntax supports arbitrary first-order predicates, so the models can be faithfully reproduced.
%
%
%We take this approach for two reasons.
%%
%First, it allows us to automatically find and verify small counter-examples in the original model.
%%
%Second, since ARM's vendor documentation currently has no specification of the mixed-size concurrency behaviour, we expect the mixed-size specification may be subject to changes. Using Alloy's automatic search for counter-examples allows rapidly iterating the ARMv8 model and cheaply re-running the search in the face of such revisions in the future.
%%

\subsection{ARMv8 Search}

%\todojp{This adapts what Wickerson et al do.
%It is basically a bounded version of what one would do in a proof of soundness of a compilation scheme.}

We are able to use these Alloy models to test that our hand-found counter-examples are real (i.e. the execution is disallowed in JavaScript but the related execution is allowed in our ARMv8 model).
In addition, following the approach of~\citet{Wickerson:2017:ACM:3009837.3009838}, we are able to use Alloy to automatically find smaller counter-examples than we were able to find manually.
Our best hand-discovered counter-example for the ARMv8 violation required 8 events and 3 byte locations; Alloy finds a counter-example with 6 events, 2 byte locations.

In this search, we are looking for counter-examples to the ARMv8
compilation scheme. Such a counter-example is an execution \textit{Exec}\textsubscript{JS} of a
JavaScript program \textit{Prog}\textsubscript{JS} that is invalid according to the JavaScript
memory model, but which corresponds to an execution \textit{Exec}\textsubscript{ARM} of a program
\textit{Prog}\textsubscript{ARM} obtained by compiling \textit{Prog}\textsubscript{JS} to ARMv8, and where \textit{Exec}\textsubscript{ARM} is allowed by
the ARMv8 concurrency model.

To this end, we follow the approach
of~\citet{Wickerson:2017:ACM:3009837.3009838}, and define a
translation relation on candidate executions. Intuitively this should
relate a JavaScript execution \textit{Exec}\textsubscript{JS} with an ARM execution \textit{Exec}\textsubscript{ARM} if \textit{Exec}\textsubscript{JS}
and \textit{Exec}\textsubscript{ARM} are executions of the programs \textit{Prog}\textsubscript{JS} and \textit{Prog}\textsubscript{ARM}, respectively,
such that \textit{Prog}\textsubscript{JS} compiles to \textit{Prog}\textsubscript{ARM}, and \textit{Exec}\textsubscript{JS} and  \textit{Exec}\textsubscript{ARM} have the same
observable behaviour. We define a translation relation, that:
\begin{itemize}

\item is compatible with the compilation scheme:\\ events in
\textit{Exec}\textsubscript{JS} arising from JavaScript accesses
are related to events in
\textit{Exec}\textsubscript{ARM} arising from the compiled ARMv8 accesses;

\item is compatible with the program structure:\\ it preserves
  $\seqb$ edges (maps JavaScript $\seqb$ edges to the matching
  $\po$ edges in ARMv8);

\item preserves the observable behaviour:\\ preserves $\rbf$ between \textit{Exec}\textsubscript{JS} and \textit{Exec}\textsubscript{ARM}.

\end{itemize}
%
% As with~\cite{Wickerson:2017:ACM:3009837.3009838}, we treat \rbf as operationally-determined, even though it is technically existentially quantified in the definition of the memory model, since it is always possible to pick distinct written values that make \rbf precisely known.
%
%
We give the event-to-event mapping of this translation below;
we omit the (unsurprising) details of the mappings on
  relations of the candidate executions here, but give the full
  definition in the supplemental material~\cite{papersuppl}.
The event mapping is one-to-one, except JavaScript
\K{RMW} events which are implemented using a pair of load/store exclusive instructions
As a minor edge-case, if the Wasm access is an unaligned non-atomic generated by a DataView, each byte of the Wasm access must be mapped to a separate single-byte ARM event of the relevant type~\cite{Flur:2017:MCA:3009837.3009839}.
% Note that location and incoming/outgoing operationally-determined
% edges are preserved.
\begin{figure}[h]
\small
\begin{tabular}{llll}
\multicolumn{2}{c}{\textbf{Instructions}} & \multicolumn{2}{c}{\textbf{Events}} \\
\textbf{JavaScript} & \textbf{ARMv8} & \textbf{JavaScript} & \textbf{ARMv8} \\
A-.load  & \texttt{ldar} & \Rsc & R\textsubscript{acq} \\
A-.store  & \texttt{stlr} & \Wsc & W\textsubscript{rel} \\
\_ = b[$k$]  & \texttt{ldr} &  \Run & R \\
b[$k$] = \_  & \texttt{str} &  \Wun & W \\
A-.exchange & \ldots \texttt{ldaxr}/\texttt{stlxr} \ldots & \RMWsc & R\textsubscript{e-a} $\tikz[  /litmus/append to tikz path search,
  kind=events,
  JavaScript,
  show event label=false,
  /litmus/every relation/.style=,]{ \coordinate (0) at (0,0); \coordinate (1) at (0.5,0); \draw (0) edge[sb, label pos=0.35] (1); }$ W\textsubscript{e-r} \\
\end{tabular}
%\vspace{-0.5\baselineskip}
\end{figure}

Our Alloy counter-example search looks for a JavaScript
candidate execution \textit{Exec}\textsubscript{JS} and an ARMv8 candidate execution \textit{Exec}\textsubscript{ARM}, both
well-formed, such that they are related by the translation relation,
and \textit{Exec}\textsubscript{ARM} is valid in ARMv8, but \textit{Exec}\textsubscript{JS} invalid in JavaScript.

% Given this, only the \tot part of the JavaScript candidate execution is considered existentially quantified.
%
%
% Note that JavaScript \K{RMW} operations are not allowed to fail, and full compilation therefore involves wrapping the paired exclusives in a loop~\cite{ccompilation}.
%
% Our relation, like those of~\citet{Wickerson:2017:ACM:3009837.3009838}, under-specifies this behaviour, effectively considering only the fragment of the compiled candidate execution where the exclusive pair eventually succeeds.

\subsection{Finding Counter-Examples}

\begin{figure}[h]
\begin{tikzpicture}[%
  /litmus/append to tikz path search,
  kind=events,
  JavaScript,
  threads baseline,
]

% Thread 0
\node[simple instructions] {
  a:W\textsubscript{SC} b[0..3]=n \\
};

% Thread 1
\node[simple instructions] {
  b:W\textsubscript{Un} b[0..3]=m \\
  c:R\textsubscript{SC} b[0..3]=n \\
};

\draw (a) edge[tot] (b)
      (b) edge[disable edge routing, tot, out=0, in=-0, looseness=2] (c)
      (c) edge[disable edge routing, rf^{-1}] (a);
            
     \path[use as bounding box] (-2,0) rectangle (5.5,0); % adjust to fit
     
\end{tikzpicture}
\hspace{2em}
\begin{tikzpicture}[%
  /litmus/append to tikz path search,
  kind=events,
  AArch64,
  threads baseline,
]

% Thread 0
\node[simple instructions] {
  a:W\textsubscript{rel} b[0..3]=n \\
};

% Thread 1
\node[simple instructions] {
  b:W\textsubscript{} b[0..3]=m \\
  c:R\textsubscript{acq} b[0..3]=n \\
};

\draw (b) edge[co, swap] (a)
      (c) edge[disable edge routing, rf^{-1}] (a);
\end{tikzpicture}
\caption{False counter-example from na\"ive search.}
\label{fig:armv8-spurious}
\end{figure}
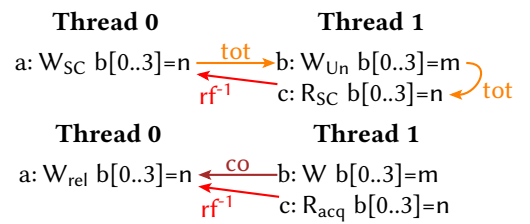

For the uncorrected JavaScript model, we would like our search to produce counter-examples similar to Fig~\ref{fig:armv8-counter}.
However, na\"ively searching as described above yields spurious counter-examples.
An example is shown in Fig~\ref{fig:armv8-spurious}.
This pair of executions satisfies the constraints of our search as specified so far: an invalid JavaScript execution, translation-related to a valid ARMv8 execution.
JavaScript here forbids the execution, because the $\rfr$ relation is
incompatible with $\totr$. However, this example is spurious, as a different choice of $\totr$ would make the execution allowed.
Any program exhibiting this candidate execution will not be a real counter-example, because it will also exhibit the candidate execution with the correct \totr, which is observably equivalent.

%!TEX root = ../main.tex

The problem illustrated by this example is due to the mismatch in the data of
ARMv8 and JavaScript candidate executions: assuming a particular ARMv8
execution, the translation relation together with the well-formedness
conditions constrains the relations of a corresponding
(translation-related) JavaScript execution, except for its
(existentially quantified) $\totr$ component. Hence the na\"ive
counter-example search will simply pick a ``bad'' \totr, that is
inconsistent with other relations of the JavaScript execution.
% Conrad: I'm not sure this paragraph gives any extra information
%
% However, note that \totr~and $\cor$ are existentially quantified, and can therefore be picked arbitrarily to satisfy the constraints.
%
%
% However, this execution is not interesting as a counter-example because the same execution can be justified simply by picking a different \totr, without changing anything else.
%
% That is, a program exhibiting this candidate execution will not be a real counter-example, because it will also exhibit the candidate execution with the correct \totr.
%
%
We are only interested in counter-examples where the JavaScript execution cannot be made valid simply by permuting \totr.
\citet{Wickerson:2017:ACM:3009837.3009838} describe
counter-example executions satisfying this requirement as having the \textit{deadness} property.\footnote{
Such executions are ``dead'' in the sense that they ``cannot move around''.}

A way of guaranteeing ``good'' counter-examples (that are \textit{dead}) would be specifying the search as the
question: ``does there \emph{exist} a valid ARMv8 execution \textit{Exec}\textsubscript{ARM}, such
that there \emph{exists} a JavaScript execution \textit{Exec}\textsubscript{JS}, that is
translation-related to \textit{Exec}\textsubscript{ARM} and such that \textit{Exec}\textsubscript{JS} is invalid in
JavaScript $\emph{for all}$ total orders $\totr$?''
%
% putting back names e and e', if we want to change, have to consistently change earlier in the section too
%
Since this Alloy search is computationally infeasible, we
use the \textit{syntactic deadness} criterion of~\citet{Wickerson:2017:ACM:3009837.3009838}.
This is a syntactic condition on candidate executions that
approximates execution deadness in a way that is computationally
feasible to check, but which may discard some legitimate
counter-examples.

For JavaScript, any condition that guarantees that candidate
executions differing only in their $\tot$ are required to
preserve \Wsc~\totr~\Wany~and \Wany~\totr~\Rsc~edges, is sufficient to
guarantee deadness (we verify this in Coq, based on the model in~\S\ref{sec:coq}).
Note in particular that the "counter-example" of Fig.~\ref{fig:armv8-spurious} does \textit{not} satisfy this condition, as the $\totr$ edge from (a) to (b) can be inverted to create a valid execution.
Defining such a search, we successfully find the counter-example in
Fig.~\ref{fig:armv8-counter} of \S\ref{sec:arm_compilation_failure}.

\subsection{Bounded Compilation Correctness}
\label{sec:alloy-lack}

With the model fixed as detailed in \S\ref{sec:deficiencies}, we
use Alloy to confirm that no counter-examples exist up to a bound (8 distinct events, 20 locations).
This also gives us the opportunity to test proof strategies in preparation for our Coq proof of compilation scheme correctness (\S\ref{sec:coq}).
In that proof, we must show that for any ARMv8-allowed execution a valid related JavaScript execution exists, which requires constructing a witnessing \totr~relation. We model checked our idea for this construction: %% and found that
making \totr~some linear extension~\cite{Szpilrajn1930}
%
% thought it was funny to cite what's now a standard result, but if you want to include it as a funny thing I'm happy to revert
%
of \mbox{$\seqbr \cup (\obsr \cap (L \cup A)^2)$}, where \mbox{$\obsr \cap (L \cup A)^2$} is ARM's observed-before relation restricted to release-acquire atomics (a full definition can be found in the supplementary appendix~\cite{papersuppl}) . With $\totr$ constrained in this way, model checking even without the syntactic deadness approximation shows the absence of compilation scheme counter examples up to the search bound.
%
% check if this is ok: I tried to hint at the fact that this is now a sound search, although only within the bounds. Did we actually model check exactly the construction we're using in the proof?

%!TEX root = ../main.tex
\begin{figure*}[h]
%\vspace{0.2\baselineskip}
%\footnoterule
%\vspace{0.8\baselineskip}
\begin{minipage}{\textwidth}
\raggedright
\figLeftMargin
% A uni-size candidate execution is considered \textit{valid} iff it satisfies all the following properties.
% \\[1em]
\begin{minipage}{0.4\textwidth}
\textbf{Happens-Before Consistency (1):}\\
$\hb \subseteq \tot$
\end{minipage}%
\begin{minipage}{0.6\textwidth}
\textbf{Happens-Before Consistency (2):}\\
$\forall \obj{E}_w \obj{E}_r \ldotp \obj{E}_w~\rf~\obj{E}_r \longrightarrow \neg (\obj{E}_r~\hb~\obj{E}_w)$
\end{minipage}
\\[1em]
\textbf{Happens-Before Consistency (3):}\\
$\forall (\obj{E}_w, \obj{E}_r) \in \rf. \quad \nexists \obj{E}'_w.~ (\obj{E}_w~\hb~\obj{E}'_w) \wedge  (\obj{E}'_w~\hb~\obj{E}_r) \wedge \textit{same-location}(\obj{E}'_w, \obj{E}_r)$
\\[1em]
\textbf{Sequentially Consistent Atomics:}\\
$\forall \obj{E}_w \obj{E}_r \ldotp
\obj{E}_w~\rf~\obj{E}_r \;\wedge\;
\obj{E}_w~\hb~\obj{E}_r
\longrightarrow$\\
$\quad \nexists \obj{E}'_w \ldotp \obj{E}'_w.\fld{ord} =  \tcons{SeqCst} \;\wedge\; \obj{E}_w~\tot~\obj{E}'_w
\;\wedge\; \obj{E}'_w~\tot~\obj{E}_r
~ \wedge$\\
$\qquad \qquad \left(
\begin{array}{r@{\,}l@{\,}l}
 & (\textit{same-location}(\obj{E}'_w, \obj{E}_r)
         &\wedge\; \obj{E}_w~\sw~\obj{E}_r )\\
\vee&
  (\textit{same-location}(\obj{E}_w, \obj{E}'_w)
         &\wedge\; \obj{E}_w.\fld{ord} = \tcons{SeqCst}
         \;\wedge\; \obj{E}'_w~\hb~\obj{E}_r )\\
\vee&
  (\textit{same-location}(\obj{E}'_w, \obj{E}_r)
         &\wedge\; \obj{E}_w~\hb~\obj{E}'_w
         \;\wedge\; ~\obj{E}_r.\fld{ord} = \tcons{SeqCst})
\end{array}\right)$
\end{minipage}
\caption{Validity of uni-size JavaScript executions.}
\label{fig:candidate-execution-valid-unisize}
%\vspace{1\baselineskip}
\end{figure*}

\subsection{SC-DRF Search}
\label{sec:sc-drf-search}
We are also able to automatically find counter-examples for SC-DRF in the uncorrected model.
We use the same search bound, and again we must use our syntactic deadness condition to remove spurious counter-examples.
We find the counter-example of Fig.~\ref{fig:sc-drf-example}, which is smaller than the hand-found counter-example of~\citet{weakeningwasm}.

\pagebreak

% For completeness, we also verify (up to a bound) the absence of counter-examples in the fixed model, although we also prove internal SC-DRF in Coq (\S\ref{sec:coq}), so this is unsurprising.

%!TEX root = main.tex
\section{Coq Verification}
\label{sec:coq}

We mechanise the JavaScript model, as shown in Figs.~\ref{fig:candidate-execution} %, \ref{fig:candidate-execution-wf}
and \ref{fig:candidate-execution-valid}, in Coq.

\subsection{SC-DRF}

We first prove that our corrected model is SC-DRF in the sense defined in \S\ref{sec:scdrf}, mechanising a previous hand-proof by \citet{weakeningwasm}:
\begin{theorem}[internal\_sc\_drf]
All well-formed, valid, data-race-free executions in the revised JavaScript model are sequentially consistent.
\end{theorem}

\subsection{Compilation Scheme Correctness}
\label{sec:coq-arm}

We now prove compilation scheme correctness, from the revised JavaScript model to our ARM model.
As mentioned in~\S\ref{sec:backgroundjs}, a limitation of this proof, not shared with our other results, is the assumption that all accesses have been generated by typed arrays (i.e. are aligned).
This simplifies the proof, since unaligned ARM accesses must be split into separate bytewise events~\cite{Flur:2017:MCA:3009837.3009839}.

We build our proof following the style of the IMM framework~\cite{Podkopaev:2019:BGP:3302515.3290382}. As in this work, the proof proceeds by defining a ``base execution'' that is shared between the two models (i.e, intra-thread program order and \rbf), and then showing that, for any such execution, validity in the ARM model implies validity in the JavaScript model.
As an intermediate lemma, we must prove that, given an allowed ARMv8 execution, it is possible to construct a witnessing \tot relation for an allowed JavaScript execution.
We achieve this proof using the construction we model-checked as part of \S\ref{sec:alloy-lack}.
The initial model-checking allowed us to rapidly validate possible constructions; it would have been far more time-consuming to come up with a correct construction from scratch.

\begin{theorem}[jsmm\_compilation]
The compilation scheme from the revised JavaScript model to (mixed-size) ARMv8 is correct.
\end{theorem}

\subsection{A Uni-Size Model}

We can define a more standard model for JavaScript assuming uni-size accesses, where disjoint byte ranges are treated as distinct abstract locations.
%
%% This is analogous to the (implicit) transformation from addresses to locations in uni-size architectural relaxed memory models.
%
% I think I don't understand this sentence. In previous papers we've used "address" and "location" interchangably. Since people have just not thought very much about mixed-size models in the past, referring to some analogy here might not really help us. But then I don't understand the sentence.
%
In the interests of space, we do not give a full definition here, but we reproduce the uni-size validity condition in Fig.~\ref{fig:candidate-execution-valid-unisize}.
It is easy to see that it is a cut-down version of Fig.~\ref{fig:candidate-execution-valid},
where references to \rbf are replaced with references to \rf, and references to byte ranges are replaced with a \textit{same-location} predicate.
The \textbf{Tear-Free Reads} condition is trivially true in the uni-size case, and can therefore be removed.
We mechanise our uni-size model, and a reduction from the mixed-size model to the uni-size one, proving that validity of mixed-size executions with no partial overlaps and no tearing (i.e. $\rfr^{-1}$ being functional) is equivalent to validity in the uni-size model.

We prove compilation scheme correctness of the uni-size model to several architectures, via the Intermediate Memory Model (IMM)~\cite{Podkopaev:2019:BGP:3302515.3290382,Moiseenko-al:CoRR19}:
\begin{theorem}[\small s\_imm\_consistent\_implies\_jsmm\_consistent]
The compilation schemes from uni-sized JavaScript to x86-TSO, POWER, RISC-V, ARMv7, and ARMv8 are correct.
\end{theorem}
As part of this, we prove that JavaScript \KK{Unord} accesses are no stronger than IMM \KK{Relaxed} accesses, and JavaScript \KK{SeqCst} accesses are no stronger than IMM \KK{SeqCst} accesses.

%!TEX root = ../main.tex

\begin{figure*}[h]
\begin{subfigure}{\textwidth}
\centering
\begin{tikzpicture}[%
  /litmus/append to tikz path search,
  kind=assem,
  JavaScript,
  threads baseline,
]
% Thread 0
\node[simple instructions] {
  |a: Atomics.wait(x, 0, 0);| \\
  |b: r0 = Atomics.load(x, 0);| \\
};

% Thread 1
\node[simple instructions] {
  |c: Atomics.store(x, 0, 42);       | \\
  |d: r1 = Atomics.notify(x, 0);       | \\
};

 \node at (2.5,1)   (x) {\texttt{x = new Int32Array(new SharedArrayBuffer[4]);}};
\end{tikzpicture}
\caption{wait/notify.}
\label{fig:wait-notify}
\end{subfigure}\\[-0.4\baselineskip]

\begin{subfigure}{0.5\textwidth}
\centering
\begin{tikzpicture}[%
  /litmus/append to tikz path search,
  kind=events,
  JavaScript,
  threads baseline,
]
% Thread 0
\node[simple instructions] {
  a1:R\textsubscript{SC} x[0..3]=0 \\
  a2:E\textsubscript{wake} x 0 \\
  b:R\textsubscript{SC} x[0..3]=0 \\
};

% Thread 1
\node[simple instructions] {
  c:W\textsubscript{SC} x[0..3]=42 \\
  d:E\textsubscript{notify} x 0=1 \\
};

 \node at (1.6,1.5)   (x) {W\textsubscript{I} x[0-3]=0};

\draw
      (x) edge[multi'={rf,hb}, swap, in=15, out=-90, label pos = 0.4] (a1)
      (d) edge[shorten <=0.1cm, asw, swap, label pos = 0.65, dashed] (a2)
      (x) edge[rf, in=0, out=-90, label pos = 0.4] (b)
      (x) edge[hb, in=165, out=-90] (c);
\end{tikzpicture}
\caption{The interleaving a $\rightarrow$ c $\rightarrow$ d $\rightarrow$ b}
\label{fig:wait-notify-exec-a}
\end{subfigure}%
\begin{subfigure}{0.5\textwidth}
\centering
\begin{tikzpicture}[%
  /litmus/append to tikz path search,
  kind=events,
  JavaScript,
  threads baseline,
]
% Thread 0
\node[simple instructions] {
 % a2:E\textsubscript{wait} \\
  a1:R\textsubscript{SC} x[0-3]=0 \\
  % b:R\textsubscript{SC} x[0-3]=0 \\
};

% Thread 1
\node[simple instructions] {
  c:W\textsubscript{SC} x[0-3]=42 \\
  d:E\textsubscript{notify} x 0=0 \\
};

 \node at (1.6,1.5)   (x) {W\textsubscript{I} x[0-3]=0};

\draw
      (x) edge[multi'={rf,hb}, swap, in=15, out=-90] (a1)
      (d) edge[shorten <=0.1cm, asw, color = asw color, swap, in=-15, out=-180, looseness=0, label pos = 0.5, dashed] (a1)
      (x) edge[hb, in=165, out=-90] (c);
      % (x) edge[rf, in=0, out=-90, label pos = 0.3] (b);
\end{tikzpicture}
\caption{The interleaving c $\rightarrow$ d $\rightarrow$ a (gets stuck)}
\label{fig:wait-notify-exec-b}
\end{subfigure}%
%\vspace{-0.5\baselineskip}
\caption{These two candidate executions for Fig.~\ref{fig:wait-notify} are forbidden if the model adds the grey edges.}
\label{fig:wait-notify-exec}
%\vspace{2\baselineskip}
\end{figure*}

\subsection{Uni-Size Programs}

Having defined our uni-size model and verified a reduction between executions of the mixed- and uni-size models, we must ask the question: under what restrictions does a JavaScript program produce only uni-size-reducible executions?
Recalling the conditions on our proof of validity-preservation, if every SharedArrayBuffer is accessed through only a single typed array, it is guaranteed that there are no partially overlapping accesses.

However, there is still the possibility of tearing.
Tearing accesses are treated as though decomposed into individual byte-wise accesses.
This means that even assuming the execution has no partial overlaps, $\rfr^{-1}$ could be non-functional (relating a read with multiple writes).
JavaScript's sequential semantics guarantees that 8, 16, and 32-bit integer typed arrays will always produce tearfree accesses.
However, even assuming that every typed array is one of these kinds, it is not guaranteed that $\rfr^{-1}$~is functional.

This is because the \textbf{Tear-Free Reads} validity rule (Fig.~\ref{fig:candidate-execution-valid}) only applies to events with identical ranges.
However, the \KK{Init} event ranges over the entire memory, and thus even fully-aligned, identically ranged tearfree accesses may observe interleaving bytes from the \KK{Init} event.
Effectively, whether we model the \KK{Init} event as tearfree or not, it will cause tearing anyway.
Consider the program of Fig.~\ref{fig:init-uni-size}.
The 16-bit load of Thread 0 is allowed by the model to read one byte from the \KK{Init} event and one byte from Thread 1's write, even though all loads and stores in the program are tearfree.

%!TEX root = ../main.tex

\begin{figure}[H]
\begin{tikzpicture}[%
  /litmus/append to tikz path search,
  kind=assem,
  JavaScript,
  threads baseline,
]
% Thread 0
\node[simple instructions] {
  |r = b[0];// r=0x0001| \\
};

% Thread 1
\node[simple instructions] {
  |b[0] = 0x0101;|\\
};

%% \draw (a.south) edge (b.south);

 \node at (1.5,1)   (x) {\texttt{b = new Uint16Array(new SharedArrayBuffer[32])}};

\end{tikzpicture}

\vspace{1em}
\begin{tikzpicture}[%
  /litmus/append to tikz path search,
  kind=events,
  JavaScript,
  threads baseline,
]
% Thread 0
\node[simple instructions] {
  a:R\textsubscript{Un} b[0..1]=0x0001 \\
};

% Thread 1
\node[simple instructions] {
  b:W\textsubscript{Un} b[0..1]=0x0101 \\
};

\node at (2.2,1)   (x) {W\textsubscript{I} b[0..31]=0};

\draw
      (x) edge[multi'={rf,hb}, swap, in=15, out=-90] (a)
      (b) edge[rf, swap,shorten <=0.1cm] (a)
      (x) edge[hb, in=165, out=-90] (b);
\end{tikzpicture}
\vspace{-0.5\baselineskip}
\caption{A tearing behaviour involving the \KK{Init} event.}
\label{fig:init-uni-size}
\vspace{-0.5\baselineskip}
\end{figure}

We believe this execution should not be allowed.
If \textbf{Tear-Free Reads} is strengthened as follows, then $\rfr^{-1}$ is guaranteed to be functional (assuming our typed array restrictions).
\begin{figure}[H]
\vspace{-0.5\baselineskip}
\begin{tabular}{@{}l}
\textbf{Tear-Free Reads (strong):}\\
$\forall \obj{E}_r \ldotp \obj{E}_r.\fld{tearfree}~\longrightarrow$\\
$\qquad\left|\left\{ 
\pbox{\textwidth}{  $\obj{E}_w~\left|~
\pbox{\textwidth}{ $\obj{E}_w~\rf~\obj{E}_r \wedge \obj{E}_w.\fld{tearfree}~\wedge$ \\ $ (\func{range}_w(\obj{E}_w) = \func{range}_r(\obj{E}_r) \vee \obj{E}_w.\fld{ord} = \ordi) $ }
\right.$  }
\right\} \right| \leq 1$
\end{tabular}
\vspace{-0.5\baselineskip}
\end{figure}

This condition intuitively seems like it should hold, and we continue to investigate whether it can be officially adopted.
%%
%With this additional strengthening, we can give a condition for $\rfr^{-1}$ to be functional.
%
%\begin{lemma}[rf\_is\_func]
%Given \obj{Exec}, a well-formed, valid execution that consists (aside from the \KK{Init} event) only of aligned, \textit{tearfree} accesses which do not partially overlap in range, if \obj{Exec} obeys \textnormal{\textbf{Tear-Free Reads (strong)}}, then $\rfr^{-1}$ is functional.
%\end{lemma}
%
Note that our uni-size compilation result applies even without the revised \textbf{Tear-Free Reads} condition: the uni-size model is a \textit{stronger} model for JavaScript programs with no partial overlaps, and nevertheless supported by the compilation schemes.

%!TEX root = main.tex
\section{Atomics.wait/Atomics.notify}
\label{sec:wait-notify}

Beyond the fragment of the language that just involves memory
accesses, JavaScript defines the thread synchronization operations \texttt{Atomics.wait} and \texttt{Atomics.notify}.
These operations are explained by way of an example program (Fig.~\ref{fig:wait-notify}).
All operations are to the SharedArrayBuffer in \texttt{x}.
The \texttt{Atomics.wait} operation reads memory location $0$, and compares the result to an expected value, $0$.
If the expected value does not match the read value, execution continues as normal.
If the expected value matches the read value, the thread suspends execution, placing itself in a \textit{wait queue} associated with the read location.
The \texttt{Atomics.notify} operation of thread 1, to the same location, will wake all threads in the wait queue for that location.
The return value of \texttt{Atomics.notify} is the number of threads woken.

Intuitively, this program should always terminate, with the load of line (b) guaranteed to read 42.
If thread 0 executes \texttt{Atomics.wait} first, it will suspend until both (c) and (d) have executed, meaning (b) will not execute until 42 has been written.
Alternatively, if thread 1 executes \texttt{Atomics.notify} first, then it will already have executed line (c) and written 42 in location 0.
Therefore thread 0's \texttt{Atomics.wait} should continue execution as it does not observe its expected value.
However, this intuition relies on the operations providing ordering guarantees, though synchronization, which are not explicitly represented in the axiomatic memory model.

Interactions with the wait queue are specified purely using an interleaving of the thread-local semantics.
The specification informally describes threads updating the wait queue as entering and leaving a lock-like ``critical section''.
However, it does not describe how the interleaved order of entry into the critical section affects the candidate executions permitted by the axiomatic memory model.
We correct this so that entering the critical section implies synchronization edges in the candidate execution to all previous exits.
This is in line with the treatment of locks in C/C++11 and the \textit{monitor lock} of the Java memory model.
It also fits the informal understanding of JavaScript implementers, who reported that they currently implement lock-like synchronization for \texttt{Atomics.wait/notify}~\cite{wait-notify-github}.

These additional synchronization edges are necessary to ensure that the axiomatic model correctly forbids intuitively disallowed executions.
Fig.~\ref{fig:wait-notify-exec-a} shows an undesirable execution where (b) reads 0 even though it cannot have executed until (d) notifies (a).
Similarly, in Fig.~\ref{fig:wait-notify-exec-b}, (a) reads 0, suspending, even if (d) records that there were no threads notified, meaning (c) must have already executed.
Incorporating the critical section entry ordering guarantees as additional-synchronizes-with edges (given by the dashed grey lines) ensures that these executions are forbidden.

%!TEX root = main.tex

\section{Related Work}
\label{sec:related}

As mentioned in~\S\ref{sec:intro-mixed-size}, there is little prior work dealing with mixed-size relaxed memory models.
\citet{Flur:2017:MCA:3009837.3009839} give mixed-size operational
models for ARMv8 and POWER; \citet{PulteEtAl2018}
adapt this model for a revision of the ARMv8 concurrency
architecture. The mixed-size ARMv8 axiomatic model presented here is
directly based on this work: generalising the ARMv8 axiomatic model
\cite{deacon-cat,PulteEtAl2018} to allow for the relaxed mixed-size
behaviours described by their operational model.
Flur et al. also describe an extension to C/C++11's model, adding mixed-size non-atomics, and give a sketch hand-proof that the resulting model can be correctly compiled to POWER.
The model of mixed-size C/C++11 is substantially simpler than JavaScript's, since non-atomics are never allowed to race.
%
% This means that, unlike JavaScript, mixed-size accesses may never interleave with each other in an unsynchronized way.
%
Moreover, our verification of JavaScript compilation is machine-checked. % , and far more rigorous.

\citet{weakeningwasm} describe the memory model of WebAssembly, and note that it is intended to be a superset (feature-wise) of JavaScript's.
The authors do not attempt verification of their proposed compilation scheme, leaving it an open problem.
The core of the WebAssembly model is inherited from JavaScript, and therefore benefits from our adopted fix.
% Our work can also be seen as a first step towards the verification of
% the WebAssembly compilation scheme. WebAssembly's memory model, however,
% contains additional behaviours such as its bounds-checking
% semantics, which we do not address here.
%
% Already say that in future work

The most closely related work to our Alloy development is that of~\citet{Wickerson:2017:ACM:3009837.3009838}.
%
% We investigated the possibility of adapting their model for this
% purpose, but found it easier to write our own: in their setting the
% axiomatic model does not need to specify well-formedness conditions
% because their litmus tests generating well-formed executions by
% construction, while we wanted to 
%
Our counter-example generation closely follows their approach for
finding compilation violations for uni-size models with Alloy.
We here extend this methodology to mixed-size models, although only for JavaScript and ARMv8 specifically, 
while their work is designed to compare arbitrary uni-size herd \cite{Alglave:2014} models, allowing them to
apply their tool to several existing models.

EMME~\cite{emmejs} is an Alloy-based tool for the (uncorrected) JavaScript memory model, 
% %
% EMME is intended
primarily intended as a test oracle.
% ; given a small program, it can calculate the permitted executions.
%
The authors identify and correct some earlier issues in the model, mainly related to well-formedness of certain definitions.
% , presumably discovered as part of the definition and execution of the model in Alloy.
%
For example, they identify that an earlier version of the model allowed RMW events to read from themselves.
Their work does not concentrate on a qualitative assessment of the model, and thus does not identify the issues we describe (\S\ref{sec:deficiencies}).
Hence, while we also use Alloy, our aim is different here.
We found it easier to write our own JavaScript model than to adapt their model to fit with Memalloy's approach.

%
% Their approach makes sense in the uni-size context, because they can leverage the various existing herd model definitions.
%
% In contrast, we develop the ARMv8 model ourselves, and there are no other existing axiomatic mixed-size models that we can use.
%
% As more mixed-size models are developed, a more general approach will be valuable.

Some core definitions of the JavaScript memory model are shared with the C/C++11 model of~\citet{Batty:2011}, but extended to a mixed-size context.
The C/C++11 model itself has been extensively formally investigated, 
%
% \citet{Nienhuis:2016} give an equivalent, incremental, operational model.
%
% Such a model would also be valuable for JavaScript as, similar to C/C++11, the axiomatic model must be defined over whole-program executions.
% \todojp{Unless you're trying to increase Kyndylan's h-index, I'm not convinced the previous sentence is very useful.}
%
% There have been many investigations into
including the correctness (or otherwise) of compilation from C/C++11~\cite{Batty:2011,Sevcik:2013:CVC:2487241.2487248,Batty:2012,Vafeiadis:2015:CCO:2676726.2676995,Lahav:2017:RSC:3062341.3062352,Podkopaev:2019:BGP:3302515.3290382,Wickerson:2017:ACM:3009837.3009838}.
Several of these works, through informing the C/C++11 compilation scheme, have influenced the compilation scheme now used by JavaScript.

%!TEX root = main.tex

\section{Future Work}
%we should keep future work

We invest significant effort into defining and validating a mixed-size
relaxed memory model for ARMv8. We benefit from the extensive body of existing work on
the ARMv8 (and the related Power) memory model.
To investigate compilation to other architectures, more work is needed to define their mixed-size behaviours.
Most glaringly, we lack a formal model of mixed-size x86, one of the most common target platforms for JavaScript.
Moreover, our ARMv8 model sidesteps some outstanding questions
about the architecture's mixed-size behaviour, by, in doubt,
choosing a reasonable weak option.
While sufficient to justify compilation correctness, more work may be needed to improve the fidelity of the model, so it is not weaker than necessary.

JavaScript will likely one day be extended with release/acquire atomics in the style of C/C++11.
We hope to engage with the standards body and use the memory model formalisation to inform such extensions.

WebAssembly's relaxed memory model is a superset (feature-wise) of JavaScript's.
%
% Our mechanised JavaScript model-internal SC-DRF proof follows the pen-and-paper proof of~\citet{weakeningwasm}, which introduces the WebAssembly memory model.
%
% That work leaves verification of WebAssembly compilation as an open problem.
%
Our approach is a first step towards verifying WebAssembly's compilation scheme, although WebAssembly's dynamic memory growth and relaxed bounds checking semantics present significant complications.
%
%We note that the web-based nature of JavaScript/WebAssembly give us a new avenue to perform litmus testing: through a web browser.
%%
%A significant limitation
% % when attempting to experimentally evaluate 
%for experimental testing of architectural memory models 
%%
%is the difficulty and expense of procuring a broad hardware testbed.
%%
%Through JavaScript/WebAssembly, it may be possible to crowdsource some litmus testing, enabling access to a wider variety of hardware than has been previously possible.
%%
%The website we constructed in~\S\ref{sec:experimental} is a prototype for this approach.

Several formalisations exist for (fragments of) JavaScript's sequential semantics~\cite{10.1145/2535838.2535876,10.1145/2737924.2737991,DBLP:conf/ecoop/GuhaSK10}.
An executable mechanisation combining JavaScript's sequential and concurrent semantics would be valuable, possibly following the approach of~\citet{10.1145/2983990.2983997} for the C/C++11 concurrency model.

\section{Conclusion}
JavaScript is a widely used language, and it is important
that its shared memory concurrency is correctly specified and
verified.
In this paper, we investigate specification
deficiencies: violations of ARMv8 compilation, and model-internal
SC-DRF.
We verify in Coq that our proposed fixes are correct, a first for a mixed-size model.
%We verify our proposed fixes in Coq, a first for a mixed-size model.
%
To that end, we develop a mixed-size ARMv8
axiomatic model.
Through collaboration with the standards committee, our
fixes will be included in future versions of the specification.

%% Acknowledgments
\begin{acks}                            %% acks environment is optional
                                        %% contents suppressed with 'anonymous'
  %% Commands \grantsponsor{<sponsorID>}{<name>}{<url>} and
  %% \grantnum[<url>]{<sponsorID>}{<number>} should be used to
  %% acknowledge financial support and will be used by metadata
  %% extraction tools.
  We thank the members of ECMA TC39 for useful discussions.
  We thank Lars T Hansen and Peter Sewell for their support and feedback. 
  This work was partly supported by the EPSRC Programme Grant
  \emph{REMS: Rigorous Engineering for Mainstream Systems}
  (EP/K008528/1). This work has received funding from the European Research Council (ERC) under the European Union's Horizon 2020 research and innovation programme (grant agreement 789108, ELVER). The first author was supported by an EPSRC DTP award
  (EP/N509620/1) and a Google PhD Fellowship in Programming
Technology and Software Engineering. The third author was supported by JetBrains Research and RFBR
  (grant number 18-01-00380). The fourth author was supported by ENS Rennes. The fifth author was supported by OCaml Labs.
\end{acks}

%% Bibliography
\bibliography{references}

%% Appendix
%\balance
%\pagebreak

%\input{appendix}

%Text of appendix \ldots

\end{document}